# Inside Perspectives on Classical Electromagnetism

Leif Pettersson

*Swedish Defence Research Agency, FOI; P.O. Box 1165, SE-581 11 Linköping, Sweden*
*e-mail: leipet@foi.se*

**Abstract**

The concept "Classical Electromagnetism" in the title of the paper here refers to a theory built on three foundations: relativity principles, the original Maxwell's equations and the mathematics of exterior calculus. In this theory of electro-magnetism, the general laws are expressed as geometric (coordinate-free) relations between quantities on a four-dimensional spacetime manifold. Based on these laws, and exterior calculus on metric spaces, we derive fields and relations that describe electromagnetism from the inside perspective of Lorentz observers, inside the Minkowski spacetime (the ordinary flat spacetime). It is shown how fields and relations defined on the spacetime are sliced into three-dimensional fields and relations on time slices, and that Lorentz observers perceive these as time-dependent fields and relations on a single three-dimensional Euclidean space. In exterior calculus (on metric spaces) each field is associated to four field variants from which one can choose how a field is to appear. Using this feature, we derive alternative equation systems describing electromagnetism from the inside perspective. In one system, A, all field quantities are form fields. In another system, C, the fields are represented as vector and scalar fields. But in contrast to the traditional Maxwell's equations with div and curl operators, there is here no need for pseudo-vectors and pseudo-scalars. Some theorems of a pseudo-free vector analysis, adapted to the C-system, are derived. The paper concludes with a section where we analyse properties of the electromagnetic field and its laws with respect to mirror transformations.

## 1. INTRODUCTION

The concept "Classical Electromagnetism" in the title of the paper refers to a theory built on three foundations: *relativity principles*, the *original Maxwell's equations* and the mathematics of *exterior calculus*.

According to the relativity principle of general covariance, the fundamental descriptions of the physical world are in the form of geometric (coordinate-free) relations between quantities on a general four-dimensional spacetime arena. This is the *outside perspective* (or "bird perspective") on physics, because the physical reality can be said to be viewed, by an imaginary observer, from outside of space and time. Mathematically, spacetime is characterized as a pseudo-Riemannian manifold and the physical quantities as tensors and tensor fields defined on the manifold.

Differential geometry is a mathematical theory for the study of pseudo-Riemannian manifolds (metric spaces) and its tensors. Exterior calculus is a theory for completely antisymmetric tensors, and because - as will also be shown in the paper - all (basic) electromagnetic quantities can be described by completely antisymmetric tensor fields, exterior calculus is ideal to use in order to mathematically describe electromagnetism. However, it is not quite enough to only use exterior calculus because the metric concept is missing in that theory. Therefore, the mathematical theory used in the paper is exterior calculus applied on metric spaces, i.e. spaces which have a metric tensor defined on them. The Appendix gives a summary of exterior calculus concepts, notations and relations that are used in the paper. References for the content of Appendix have been [7] and [8] in the reference list, which also refers to some papers, and two books, dealing with electromagnetic theory in differential-form formalism.

The outside perspective gives a complete overview of electromagnetism for all kinds of spacetimes. From this perspective, the laws are geometric (coordinate-free) relations between field quantities on a general four-dimensional spacetime manifold. Electromagnetic fields defined on spacetime will be denoted by the letter F, and its source fields by J. (Later on we will also include a source field $J^M$, representing magnetic charges and currents.) A field, of some kind, is a quantity that is defined in all points of a space, and as is shown further on, both F and J have an own set of four field variants from which one can choose how the fields are to appear. The laws can be obtained by expressing (selected parts of) Maxwell's original equations in geometric form on the spacetime manifold. The general laws are thus a generalization of Maxwell's original equations into four-dimensional spacetime, brought about by relativity principles and the mathematics of exterior calculus.

Based on these laws, and by using exterior calculus, we derive fields and relations that describe electromagnetism from the *inside perspective* of Lorentz observers, situated inside the Minkowski spacetime (i.e. the ordinary flat spacetime). Basic to the derivation is a slicing of the spacetime manifold, $M$, into a continuous set, $\{m_{x^0}\}$, of three-dimensional subspaces, called time slices. Each slice represents the ordinary Euclidean space at a certain Lorentz time $x^0 = ct$, as perceived by observers stationary relative the system $\{m_{x^0}\}$. ($c$ denotes the velocity of light.) For other Lorentz observers, moving with constant relative velocity, spacetime is sliced into another set, $\{m_{x^0}\}$, related to $\{m_{x^0}\}$ by Lorentz transformations.

Instead of a single spacetime manifold on which the electromagnetic field F and its source field J are defined, we have a set of three-dimensional spaces on which one must find three-dimensional fields that can represent the four-dimensional F and J. In section 3 is shown how F can be decomposed into two sets, which we denote $\{E_{x^0}\}$ and $\{H_{x^0}\}$, and where $E_{x^0}$ and $H_{x^0}$ are three-dimensional fields defined on slice $m_{x^0}$. The source field J is sliced into two sets denoted $\{c\rho_{x^0}\}$ and $\{j_{x^0}\}$. The fields in the sets are defined on separate spaces $m_{x^0}$, but which have identical metrical properties. Therefore, as will be shown, all fields can be "mathematically transferred" to a single Euclidean space by a straightforward procedure. Considered in this way, the four-dimensional fields F and J have been decomposed into four sequences of three-dimensional fields on one Euclidean space.

Seen from an outside perspective, stationary Lorentz observers relative the system $\{m_{x^0}\}$ traverse spacetime along $x^0$-coordinate lines, which are their world lines, and cross the slices with the fields $E_{x^0}$, $H_{x^0}$, $c\rho_{x^0}$, $j_{x^0}$. According to the standard view, this is perceived by the observers as a passage of the time $x^0$ in a single three-dimensional Euclidean space, and where things change with this time. Thus, from the observers inside perspective, the sliced parts of the electromagnetic field on the different slices are described as two time-varying fields, $E_{x^0}$, $H_{x^0}$, on one three-dimensional Euclidean space. In a corresponding way, the source field J is described as the time-varying fields $c\rho_{x^0}$ and $j_{x^0}$ on the same space. These four time-dependent fields correspond to just mentioned "mathematically transferred" sequences.

Up to now the field quantities F, J, $E_{x^0}$, $H_{x^0}$, $c\rho_{x^0}$, $j_{x^0}$ have not been assigned a field type. An important aspect in the paper is that to each of these six quantities there are four field variants from which one can choose how the fields are to appear. The fields of exterior calculus are p-form fields and p-vector fields, i.e. completely anti-symmetric tensor fields of type $(0/p)$ and $(p/0)$, respectively. The rank $p$ can take on the integer values 0 up to $n$, where $n$ is the dimension of the space on which the field is defined. 0-vector field and 0-form field are the same as scalar field, and 1-vector field the same as vector field. Due to the antisymmetry, the number of independent components for a p-form or p-vector is equal to $n!/(p!(n-p)!)$. It follows from this formula that p-vectors, p-forms, (n-p)-forms and (n-p)-vectors have the same number of independent components, and therefore one can find four one-to-one mappings between these variants of field. The mappings are of two kinds: index raising and index lowering transformations and two sorts of dual transformations, and they are defined from the metric, including the volume measure, of the manifold (see Appendix). Index raising and index lowering operators map p-forms into p-vectors and vice versa. Dual operators map p-forms into (n-p)-vectors and p-vectors into (n-p)-forms.

Consequently, each field is associated to a set of four field variants. For example, the four-dimensional electromagnetic field F can appear as a 2-form field $\widetilde{F}$, as another 2-form field $\overset{\ast}{\widetilde{F}}$ (where the star is a label and not an operator), as a 2-vector field $\overline{F}$ and as another 2-vector field $\overset{\ast}{\overline{F}}$. (These types of field have six independent components.) A tilde over a letter denotes a p-form of some rank, and a bar over a letter indicates a p-vector. Thus, with these conventions, which are one standard, one can not tell the rank of the p-form or p-vector. On three-dimensional spaces we will use the convention that 2-forms are indicated by double tilde signs and 2-vectors by double bar signs. A single tilde will therefore (on three-dimensional spaces) indicate a 1-form, apart from the volume-form $\widetilde{\omega}$ (which is a 3-form), and a single bar will indicate a vector (1-vector), apart from the 3-vector $\overline{\omega}$. As an example of field variants on the three-dimensional Euclidean space, the $E_{x^0}$-field can appear as a 1-form field $\widetilde{E}_{x^0}$, as a vector field $\overline{E}_{x^0}$, as a 2-form field $\widetilde{\widetilde{E}}_{x^0}$ or as a 2-vector field $\overline{\overline{E}}_{x^0}$. (These types of field have three independent components.)

It can be pointed out here that we consider tensors as linear operators which operate on vectors and 1-forms. For example, the 2-form field $\widetilde{\widetilde{E}}_{x^0}$, or more detailed $\widetilde{\widetilde{E}}_{x^0}(\ ,\ )$, is thus an operator field with "slots" for two vector arguments. Operating with $\widetilde{\widetilde{E}}_{x^0}$ on two small vectors gives the E-field flux through the parallelogram formed by the vectors. Because $\widetilde{\widetilde{E}}_{x^0}$ is a 2-form, and thus an antisymmetric tensor, the flux changes sign on interchange of the vector arguments.

In section 3A, the electromagnetic field represented as the 2-form field $\widetilde{F}$ is sliced. The result is two sets with three-dimensional fields, $\{\widetilde{E}_{x^0}\}$ and $\{\widetilde{\widetilde{H}}_{x^0}\}$, i.e. fields represented as 1-form and 2-form. If the electromagnetic field is represented as the other 2-form, $\overset{\ast}{\widetilde{F}}$, the sliced parts are $\{-\widetilde{H}_{x^0}\}$ and $\{\widetilde{\widetilde{E}}_{x^0}\}$.

In section 3B, the general electromagnetic laws, $\mathbf{d}\widetilde{F} = 0$, $\mathbf{d}\overset{\ast}{\widetilde{F}} = \overset{\ast}{}\widetilde{J}$, are sliced. This will to lead to four time-dependent geometric relations between quantities defined on the ordinary Euclidean space. (Here geometric means coordinate-free with respect to that space.) See the A-system and the general laws in Figure 1. $\mathbf{d}$ denotes the exterior derivative and $\ast$ a dual operator (see Appendix for definitions). The unit of F, $E_{x^0}$, $H_{x^0}$ is Ampere/m, and the unit of J, $j_{x^0}$, $c\rho_{x^0}$ is Ampere/m². The A-system is consequently an inside perspective on general laws of electromagnetism as perceived by Lorentz observers situated inside the Minkowski spacetime. (From the outside perspective, the laws of electromagnetism appear as static geometric relations on the four-dimensional spacetime, but from the inside perspective they appear as dynamic geometric relations on the Euclidean space.) Fields represented as 1-form and 2-form in the A-system correspond to Maxwell's two types of vector quantities, "intensity vectors" and "flux vectors". Here the two variants of field, the 1-form field and the 2-form field, follow automatically in the derivation of the A-system. An explanation to the connections between Maxwell's two types of vector and the 1-form and 2-form is given in section 3B.

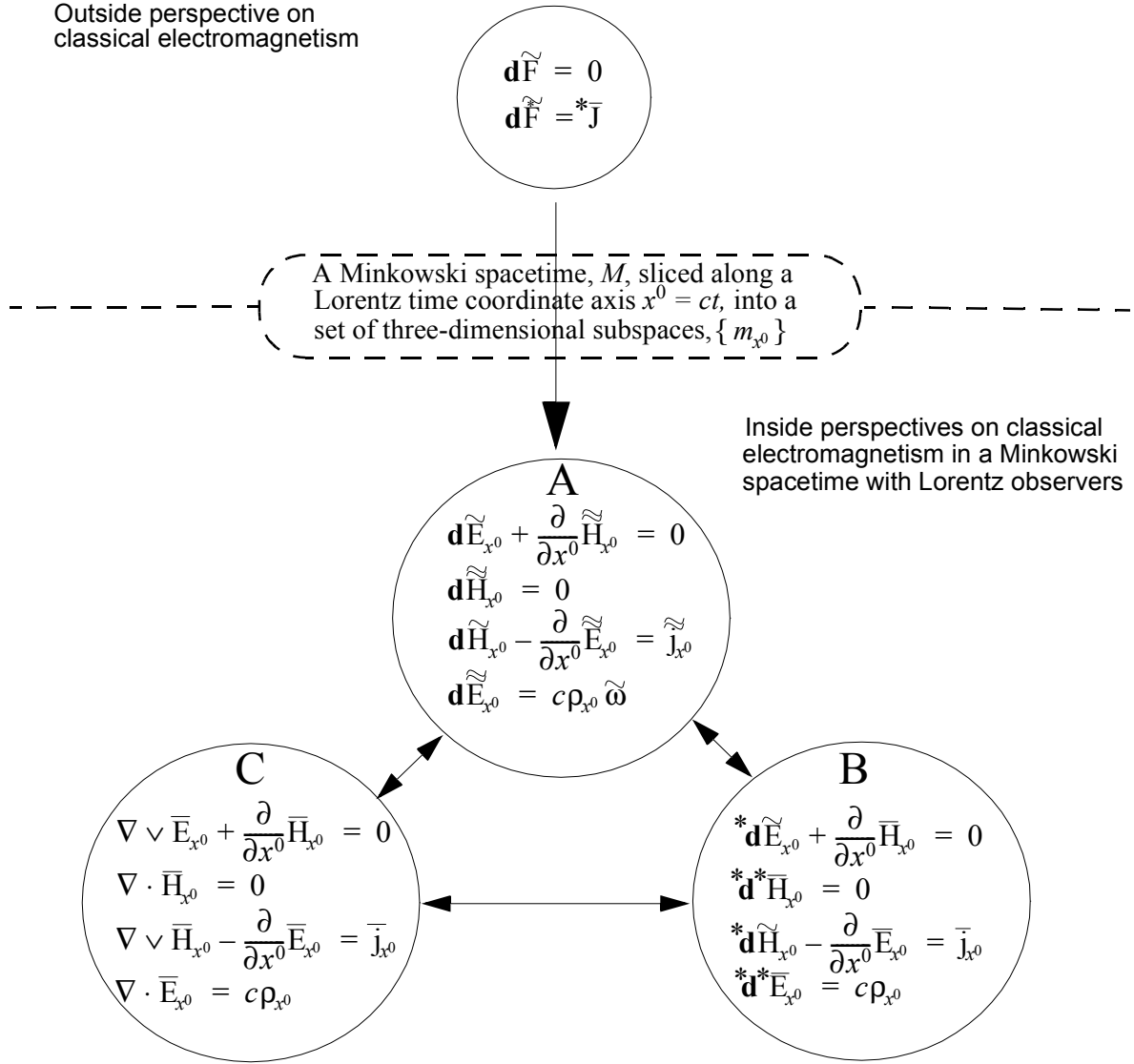

Figure 1. Outside and inside perspectives on classical electromagnetism.

Because we have four options to represent a field, the A-system can be transformed - with the aid of index raising or index lowering operators and two variants of dual operators - into other systems where the fields are represented in alternative ways, see Figure 1. An index raising operator is denoted $R^{op}$, an index lowering operator $(R^{op})^{-1}$ and dual operators $*$; $\mathbf{g}$ is the metric tensor of the Euclidean space. In the C-system, the fields are represented as vector fields and scalar fields. That is similar to the traditional Maxwell's equations, with div and curl operators, but in the C-system there is no need for introducing the concepts pseudo-vector and pseudo-scalar as in the traditional Maxwell's equations. The reason pseudo-quantities must be used is an insufficient definition of the curl operator (and also of the cross product). When the curl operator operates on a vector field the result is a pseudo-vector field, which can have different coordinate transformation properties than a vector field. This is in contrast to the corresponding $\nabla \vee$ operator in the C-system, see Figure 1, which transforms vector fields into vector fields. $\nabla \vee$ is a derived operator, composed of basic operators, where $(R^{op})^{-1}$ transforms a vector field into a 1-form field, $\mathbf{d}$ transforms the 1-form field into a 2-form field and $^*$ transforms the 2-form field into a vector field. Some theorems of a pseudo-free vector analysis, adapted to the C-system, are derived in section 3C. The operators of this vector analysis are composed of exterior calculus operators, and the metric tensor, and correspond to the operators div, curl, grad, cross product and scalar product of the traditional vector analysis.

The C-system is derived from general four-dimensional electromagnetic laws, it has no pseudo-quantities and it is a part of a unified theory of classical electromagnetism, schematically described by Figure 1. Therefore, it is a more basic (and rigorous) system compared to the traditional Maxwell's equations with div and curl operators.

Although no magnetic charges and currents have been found in nature, this type of source for an electromagnetic field is sometimes used as a pure mathematical tool, e.g. in antenna theory. The general laws with a magnetic source field $J^M$ included are $\mathbf{d}\tilde{F} = {}^* \tilde{J}^M$, $\mathbf{d}\tilde{F} = {}^* \tilde{J}$. In section 3D it is shown how the systems A, B, C in Figure 1 are modified if the sources of the electromagnetic field also include magnetic charges and currents.

The paper concludes with a section where we analyse properties of the electromagnetic field and its laws with respect to mirror transformations.

## 2. OUTSIDE PERSPECTIVE ON CLASSICAL ELECTROMAGNETISM

Seen from the outside perspective, a particle's "journey" through spacetime is described by its world line. This line can be characterized by a continuous set of spacetime points (events), denoted $P(\tau)$, where the parameter $\tau$ is the scalar quantity proper time. Expressed in a coordinate system $\{x^\alpha\}$ the line is given by four functions $l^\alpha(\tau)$, according to the relations

$$x^\alpha = l^\alpha(\tau), \ \alpha = 0, 1, 2, 3 \ .$$

Here we have used the common praxis to denote indices for quantities in four-dimensional spacetime with Greek letters, which run from zero to three and where index zero denotes the time coordinate. The tangent vector to the world line is $d/d\tau$, which expressed in the coordinate vector basis $\{\partial/(\partial x^\alpha)\}$ satisfies

$$\frac{d}{d\tau} = \frac{dl^\alpha(\tau)}{d\tau} \frac{\partial}{\partial x^\alpha}.$$

(Here the summation convention has been used, see Appendix.)

The inner product (scalar product) between two vectors, $\overline{A}$ and $\overline{B}$, is defined by

$$\overline{A} \cdot \overline{B} = \mathbf{G}(\overline{A}, \overline{B}), \tag{1}$$

where $\mathbf{G}( \ , \ )$ is the metric tensor of spacetime, in shorthand notation $\mathbf{G}$. (We thus consider tensors as linear operators acting on vectors and 1-forms, see Appendix.) $\mathbf{G}$ is a tensor of a general spacetime manifold with signature $(-,+,+,+)$. Locally the metric satisfies the Minkowski metric (see next section). Because $\tau$ is the proper time, the tangent vector $d/d\tau$ is the 4-velocity of the particle and will be denoted $\overline{U}$. Using one of the basic relations of relativity theory, we have $\overline{U} \cdot \overline{U} = -c^2$, i.e. seen from an outside perspective, all particles "travel" through spacetime with the same magnitude of their 4-velocities, the velocity of light $c$.

A classical electromagnetic field may be described as an "agent" which operates (acts) on electrically charged particles, characterized by their charge Q, rest mass $m_0$, and 4-velocity $\overline{U}$, causing the particles to change their states of motion. Tensor fields are linear operators, defined on each point of a space, and which can operate on vectors and 1-forms. For example, a (0/2)-tensor $\mathbf{A}( \ , \ )$ can operate on two vector arguments resulting in real numbers. If it operates on a single vector, one "slot" is left open, which means that the result of $\mathbf{A}$ operating on a vector is a (0/1)-tensor, i.e. a 1-form. It is then plausible that the action of an electromagnetic field on a charged particle can be described by the relation

$$\frac{d(m_0\tilde{U})}{d\tau} = \mathbf{A}( \ , Q\overline{U}), \tag{2}$$

where $\tilde{U}$ is the 4-velocity represented as a 1-form. The transformation from the 4-velocity vector is performed by the metric tensor, and we have $\tilde{U} = \mathbf{G}( \ , \overline{U})$. (Because $\tau$ is a scalar $d(m_0\tilde{U})/d\tau$ is a 1-form.) The relation (2) is Lorentz force law from the outside perspective, and can be said to both define an electromagnetic field, from the action it causes, and predict the action of a field. If both sides in the relation operate on $\overline{U}$, and we use the well established assumption that the rest mass is not affected by the electromagnetic field, we get

$$m_0\frac{d\tilde{U}}{d\tau}(\overline{U}) = \frac{m_0}{2}\frac{d(\overline{U} \cdot \overline{U})}{d\tau} = \frac{m_0}{2}\frac{d(-c^2)}{d\tau} = 0 = Q\mathbf{A}(\overline{U}, \overline{U}) \ .$$

Hence, $\mathbf{A}(\overline{U}, \overline{U})$ is zero for arbitrary velocity vectors $\overline{U}$ (where $\overline{U} \cdot \overline{U} = -c^2$). We also have $\mathbf{A}(\overline{b}, \overline{b}) = 0$ for arbitrary timelike vectors $\overline{b}$ (i.e. vectors with $\overline{b} \cdot \overline{b} < 0$), because $\overline{b}$ can be written as a real number times a velocity vector. It then can be show that $\mathbf{A}$ must be an antisymmetric (0/2)-tensor, i.e. $\mathbf{A}(\overline{b}, \overline{d}) = -\mathbf{A}(\overline{d}, \overline{b})$ for arbitrary vectors $\overline{b}$ and $\overline{d}$. This means that $\mathbf{A}$ is a 2-form. Because we will use mks units and want the unit of the electromagnetic field to be Ampere/m, we write $\mathbf{A} = \mu_0 \overset{\sim}{F}$ where $\mu_0$ is the permeability of free space and the electromagnetic field is represented as the 2-form field $\overset{\sim}{F}$.

Hence, the mathematics of exterior calculus can be used to describe the electromagnetic field. Every field of exterior calculus, on metric spaces, has four field variants from which one can choose how the field is to appear. The transformations between the different variants are performed by index raising and index lowering operators, denoted $R^{op}$ and $(R^{op})^{-1}$ respectively, and two kinds of dual operators, both denoted by a star, see Appendix for definitions. Figure 2 shows the different variants of field, and the transformations between these, for an electromagnetic field F. Horizontal arrows indicate index raising or index lowering operators and vertical arrows indicate dual operators. $\overline{F}$ denotes the 2-vector obtained by an index raising operation on the 2-form $\overset{\sim}{F}$, and $\overset{\sim}{\overline{F}}$ is the 2-form obtained from a dual operation on $\overline{F}$. (The star over F is a label, to distinguish this 2-form from $\overset{\sim}{F}$, and does not indicate an operator.) Formally, we have the relations

$$\overline{F} = R^{op} \overset{\sim}{F}, \; \overset{\sim}{\overset{*}{F}} = {}^* \overline{F} \;\;, \;\; \overset{*}{\overline{F}} = R^{op} \overset{\sim}{\overset{*}{F}} \;, \text{ which expressed in components are:} \tag{3}$$

$$F^{\alpha\beta} = G^{\alpha\gamma} G^{\beta\delta} F_{\gamma\delta} \;\;, \;\; \overset{*}{F}_{\alpha\beta} = \frac{\Omega_{\gamma\delta\alpha\beta} F^{\gamma\delta}}{2} \;\;, \;\; \overset{*}{F}^{\alpha\beta} = G^{\alpha\gamma} G^{\beta\delta} \overset{*}{F}_{\gamma\delta} \;. \tag{4}$$

Here index and summation conventions have been used; $G^{\alpha\gamma}$ are components of the inverse metric tensor and $\Omega_{\gamma\delta\alpha\beta}$ components of the volume-form, see Appendix.

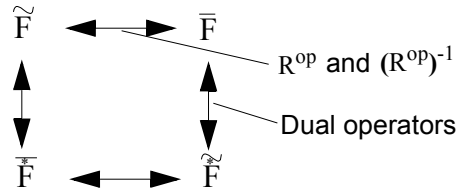

Figure 2. Field variants for the electromagnetic field.

Thus, from the outside perspective, there are two 2-form fields and two 2-vector fields that each can describe an electromagnetic field. (All the fields have six independent components.)

The exterior derivative, denoted $\mathbf{d}$, is an operator that is defined to operate on a form field of arbitrary rank, transforming it into a field one rank higher, see Appendix for a definition. Consequently, $\mathbf{d}\overset{\sim}{F}$ and $\mathbf{d}\overset{\sim}{\overset{*}{F}}$ are two 3-form fields (which have four independent components). By applying dual operators on these fields, we transform them into vector fields which we denote $\overline{J}^{M}$ and $\overline{J}$, respectively:

$$^*\mathbf{d}\overset{\sim}{F} = \overline{J}^{M} \;, \;\; ^*\mathbf{d}\overset{\sim}{\overset{*}{F}} = \overline{J}. \tag{5}$$

As will be shown - using the principle of general covariance and Maxwell's original equations - the relations (5) are general laws of classical electromagnetism, relating the electromagnetic field to its source fields. J denotes the electric (four-dimensional) current density and $J^{M}$ the magnetic current density, which we now will set to zero, as no magnetic charges have been found in nature. However, magnetic currents are sometimes used as a pure mathematical tool, e.g. in antenna theory. In section 3D it is shown how derived equation systems are modified when the sources also include magnetic charges and currents.

Because a vector field on a four-dimensional space has four components, the general laws in (5) can be separated into eight relations containing the field components of the electromagnetic field and its source field. We now consider the ordinary spacetime (with Minkowski metric) in which we introduce Lorentz-coordinates, i.e.

$$x^0 = ct, \; x^1 = x, \; x^2 = y, \; x^3 = z. \tag{6}$$

The components of the 2-form $\overset{\sim}{F}$ on the coordinate vector basis $\{\partial/\partial x^\alpha\}$, $F_{\alpha\beta} = \overset{\sim}{F}(\partial/\partial x^\alpha, \partial/\partial x^\beta)$, can be described by the component matrix in Figure 3A. The reason for denoting the components in this way is given below. The components of $\overset{\sim}{\overset{*}{F}}$ are obtained from $F_{\alpha\beta}$ with the aid of (4), where now the components of the metric tensor (and also its inverse) are given by the diagonal matrix diag(-1,1,1,1), and the components of the volume-form are given by the Levi-Civita symbol $\varepsilon_{\alpha\beta\gamma\delta}$. (This symbol is +1 if $\alpha\beta\gamma\delta$ is an even permutation of 0123, $-1$ if it is an odd permutation and zero in other cases.) The resulting component matrix is shown in Figure 3B.

$$A \qquad F_{\alpha\beta} = \begin{bmatrix} 0 & -X & -Y & -Z \\ X & 0 & N & -M \\ Y & -N & 0 & L \\ Z & M & -L & 0 \end{bmatrix} \begin{matrix} \rightarrow \beta \end{matrix} \qquad B \qquad \overset{*}{F}_{\alpha\beta} = \begin{bmatrix} 0 & L & M & N \\ -L & 0 & Z & -Y \\ -M & -Z & 0 & X \\ -N & Y & -X & 0 \end{bmatrix}$$

Figure 3. Two component matrices for an electromagnetic field.

As follows from the definition of the exterior derivative in Appendix, the components of the 3-forms $\mathbf{d}\tilde{\tilde{F}}$ and $\mathbf{d}\overset{*}{\tilde{\tilde{F}}}$ satisfy:

$$(\mathbf{d}\tilde{\tilde{F}})_{\alpha\beta\gamma} = F_{\alpha\beta,\gamma} + F_{\gamma\alpha,\beta} + F_{\beta\gamma,\alpha} \;,\; (\mathbf{d}\overset{*}{\tilde{\tilde{F}}})_{\alpha\beta\gamma} = \overset{*}{F}_{\alpha\beta,\gamma} + \overset{*}{F}_{\gamma\alpha,\beta} + \overset{*}{F}_{\beta\gamma,\alpha} \;, \tag{7}$$

where the comma notation for the partial derivatives, $\partial F_{\alpha\beta}/\partial x^\gamma = F_{\alpha\beta,\gamma}$, has been used. According to relation (5), the duals of the 3-forms in (7) give the vectors $\tilde{\mathbf{J}}^M$ and $\tilde{\mathbf{J}}$. If we set $\tilde{\mathbf{J}}^M$ to zero we obtain the following eight component relations (see Appendix for definitions of the dual operators):

$$-\frac{\varepsilon^{\alpha\beta\gamma\delta}}{3!}(F_{\alpha\beta,\gamma} + F_{\gamma\alpha,\beta} + F_{\beta\gamma,\alpha}) = 0 \;,\; -\frac{\varepsilon^{\alpha\beta\gamma\delta}}{3!}(\overset{*}{F}_{\alpha\beta,\gamma} + \overset{*}{F}_{\gamma\alpha,\beta} + \overset{*}{F}_{\beta\gamma,\alpha}) = J^\delta \;. \tag{8}$$

(The Levi-Civita symbol $\varepsilon^{\alpha\beta\gamma\delta}$ is defined in the same way as the recently defined symbol $\varepsilon_{\alpha\beta\gamma\delta}$.) By permuting indices, (8) is reduced to

$$\frac{\varepsilon^{\delta\alpha\beta\gamma}}{2}F_{\alpha\beta,\gamma} = 0 \;,\; \frac{\varepsilon^{\delta\alpha\beta\gamma}}{2}\overset{*}{F}_{\alpha\beta,\gamma} = J^\delta \;. \tag{9}$$

Substituting the component matrices of Figure 3 into (9), observing relation (6) and the comma notation for the partial derivatives, yields the following eight connections:

$$\partial L/\partial x + \partial M/\partial y + \partial N/\partial z = 0 \;,\; \partial X/\partial x + \partial Y/\partial y + \partial Z/\partial z = J^0 \;,$$

$$-\partial L/\partial(ct) - \partial Z/\partial y + \partial Y/\partial z = 0 \;,\; -\partial X/\partial(ct) + \partial N/\partial y - \partial M/\partial z = J^1 \;,$$

$$-\partial M/\partial(ct) + \partial Z/\partial x - \partial X/\partial z = 0 \;,\; -\partial Y/\partial(ct) - \partial N/\partial x + \partial L/\partial z = J^2 \;,$$

$$-\partial N/\partial(ct) - \partial Y/\partial x + \partial X/\partial y = 0 \;,\; -\partial Z/\partial(ct) + \partial M/\partial x - \partial L/\partial y = J^3 \;.$$

These relations are recognized as (one of several variants of) Maxwell's equations. For example, the last six equations, with the current equal to zero, are identical to the relations used by A. Einstein in his paper of 1905, "On the Electrodynamics of moving Bodies", where the quantities $X$, $Y$, $Z$ are described as the components of the electric force vector and $L$, $M$, $N$ as the components of the magnetic force vector. If a source field is included then $J^0 = c\rho$, where $\rho$ is the charge density, and $J^1$, $J^2$, $J^3$ are the components of a three-dimensional current density vector. But none of these vectors are relevant when characterizing electromagnetism from the outside perspective. In the next section we derive some equation systems describing the laws of electromagnetism from the inside perspective, and in one of the systems the laws are relations between three-dimensional vector fields and scalar fields.

According to the principle of general covariance, in the most general and fundamental description of the physical world, the laws are geometric (coordinate-free) relations between quantities defined on a four-dimensional spacetime. In this section we started from two such coordinate-free relations, $^*\mathbf{d}\tilde{\tilde{F}} = 0$, $^*\mathbf{d}\overset{*}{\tilde{\tilde{F}}} = \tilde{\mathbf{J}}$, where fields and operators are defined on a general four-dimensional spacetime. It was shown that for a Minkowski spacetime with Lorentz coordinates, these relations lead to eight connections between four-dimensional field components, and that the connections are (a variant of) Maxwell's original equations. Hence, using the principle of general covariance, it follows that $^*\mathbf{d}\tilde{\tilde{F}} = 0$, $^*\mathbf{d}\overset{*}{\tilde{\tilde{F}}} = \tilde{\mathbf{J}}$ are general laws of classical electromagnetism. If we apply a dual operation on both sides of these relations, and use relation (71) in Appendix, we can also express the laws as

$$\mathbf{d}\tilde{\tilde{F}} = 0 \;,\; \mathbf{d}\overset{*}{\tilde{\tilde{F}}} = {}^*\tilde{\mathbf{J}} \;. \tag{10}$$

## 3. INSIDE PERSPECTIVES ON CLASSICAL ELECTROMAGNETISM

### A. Slicing of Minkowski spacetime and its fields

Mathematically, a Minkowski spacetime is defined as a metric manifold (metric space) in which it is possible to introduce global coordinate systems, $\{x^\alpha\}$, where the components of the metric tensor on the basis $\{\partial/\partial x^\alpha\}$ satisfy $G_{\alpha\beta} = \mathbf{G}(\partial/\partial x^\alpha, \partial/\partial x^\beta) = \text{diag}(-1,1,1,1)$. The Lorentz systems, $x^0 = ct$, $x^1 = x$, $x^2 = y$, $x^3 = z$, where $x$, $y$, $z$ are Cartesian coordinates extended into four-dimensional spacetime, satisfy this criteria. Thus, the metric tensor of the Minkowski spacetime is given by

$$\mathbf{G} = -\mathbf{d}x^0 \otimes \mathbf{d}x^0 + \mathbf{d}x \otimes \mathbf{d}x + \mathbf{d}y \otimes \mathbf{d}y + \mathbf{d}z \otimes \mathbf{d}z \,, \tag{11}$$

where the outer product and the duals basis $\{\mathbf{d}x^0, \mathbf{d}x, \mathbf{d}y, \mathbf{d}z\}$ are defined in Appendix. Figure 4 shows how the extension of the Cartesian system can be illustrated for a case with the $z$-dimension suppressed. Here the $x$- and $y$-coordinate lines of the ordinary space are parallel transported along the $x^0$-coordinate lines, which are the world lines of stationary Lorentz observers (relative the system $x^0, x, y, z$), resulting in a three-dimensional coordinate grid. In a corresponding way we obtain a coordinate grid for Minkowski spacetime from the three-dimensional Cartesian coordinates $x$, $y$, $z$. Therefore, the coordinate vector basis $\{\partial/\partial x^0, \partial/\partial x, \partial/\partial y, \partial/\partial z\}$ forms a four-dimensional vector basis, and the dual basis $\{\mathbf{d}x^0, \mathbf{d}x, \mathbf{d}y, \mathbf{d}z\}$ a four-dimensional 1-form basis. The notations $\partial/\partial x$, $\partial/\partial y$, $\partial/\partial z$, $\mathbf{d}x$, $\mathbf{d}y$, $\mathbf{d}z$ can consequently mean three-dimensional vector and 1-form fields defined on the ordinary Euclidean space, or four-dimensional fields defined on the Minkowski spacetime.

The parallel transported coordinate grids span three-dimensional subspaces, called time slices, denoted $m_{x^0}$ and representing the ordinary space at the Lorentz time $x^0 = ct$. See Figure 4 where the shaded planes give a two-dimensional illustration of some three-dimensional time slices. The Minkowski spacetime, $M$, can be considered as composed of an infinite set of such slices, $M = \{m_{x^0}\}$, and conversely, spacetime can be sliced into such a set. We note that this way of slicing spacetime is connected to Lorentz observers whose world lines are the $x^0$-coordinate lines. Other Lorentz observers (in constant relative motion) follow other straight world lines which can be used to define alternative time-coordinate lines, and other sets of time slices, $\{m_{x^0'}\}$, related to $\{m_{x^0}\}$ by Lorentz transformations. (It is also possible to slice the Minkowski spacetime into curved slices which are related to accelerating observers. But in this paper such cases are not considered.)

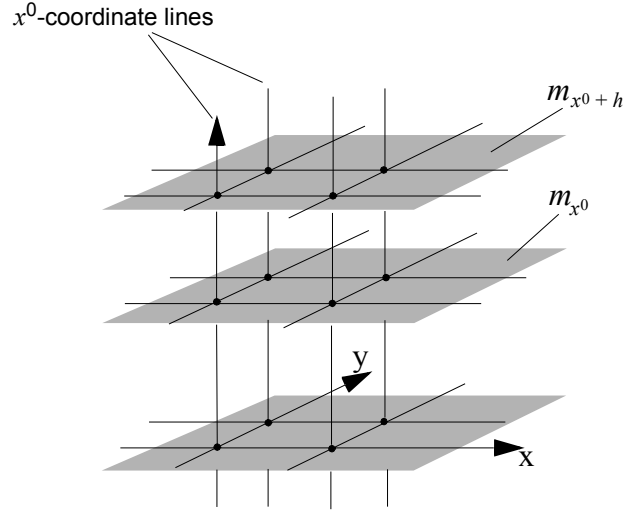

Figure 4. Illustration of time slices and coordinate lines in Minkowski spacetime.

Instead of the Cartesian coordinate system, we can use arbitrary curvilinear coordinate systems, $\{x^i, i = 1, 2, 3\}$, in the Euclidean space. In a corresponding way as above also these systems are possible to extended in order to cover the Minkowski spacetime. In this paper we will only use such coordinate systems, which thus are independent of the $x^0$-coordinate. Therefore, $x^i$ denotes a curvilinear coordinate in the ordinary space, or its extension into the Minkowski spacetime. $\partial/\partial x^i$ either means a three-dimensional vector field defined on the ordinary Euclidean space or the corresponding four-dimensional vector field on the Minkowski spacetime. (The field is independent of $x^0$.) In a corresponding way, $\mathbf{d}x^i$ denotes a 1-form field on the Euclidean space or its extended version on the Minkowski spacetime. A function $f(x^1, x^2, x^3)$ on the Euclidean space is (straightforwardly) extended into a time-independent function on the Minkowski spacetime according to $f(x^0, x^1, x^2, x^3) = f(x^1, x^2, x^3)$. A general vector field $\overline{A}$ on the Euclidean space can be written as a linear combination of the vectors $\partial/\partial x$, $\partial/\partial y$, $\partial/\partial z$, where the coefficients of the combination are functions of $x$, $y$, $z$. Extending the coefficients and the vectors yields a time-independent field $\overline{A}$ that is defined on the Minkowski spacetime. From (11) follows that $\mathbf{G}(\partial/\partial x^0, \overline{A}) = 0$, i.e. $\overline{A}$ is orthogonal to $\partial/\partial x^0$. We also have

$$\mathbf{G}\left(\frac{\partial}{\partial x^0}, \frac{\partial}{\partial x^i}\right) = 0 \quad , \quad \mathbf{G}\left(\frac{\partial}{\partial x^0}, \frac{\partial}{\partial x^0}\right) = -1 \quad , \quad G_{ij} = \mathbf{G}\left(\frac{\partial}{\partial x^i}, \frac{\partial}{\partial x^j}\right) = \mathbf{g}\left(\frac{\partial}{\partial x^i}, \frac{\partial}{\partial x^j}\right) = g_{ij} \,, \tag{12}$$

where **g** is the metric tensor of the Euclidean space,

$$\mathbf{g} = \mathbf{d}x \otimes \mathbf{d}x + \mathbf{d}y \otimes \mathbf{d}y + \mathbf{d}z \otimes \mathbf{d}z . \tag{13}$$

For the inverse metric tensors (see Appendix) we have the component relation

$$G^{ij} = g^{ij} . \tag{14}$$

We now choose one of the earlier defined extended curvilinear systems, $\{x^i\}$, together with the time coordinate $x^0 = ct$, as coordinates in the Minkowski spacetime. The electromagnetic 2-form $\widetilde{\mathrm{F}}$ is expanded according to (see the section on wedge product in Appendix):

$$\widetilde{\mathrm{F}} = F_{|\alpha\beta|} \, \mathbf{d}x^\alpha \wedge \mathbf{d}x^\beta , \tag{15}$$

where the bar summation convention has been used, i.e. the sum is over indices $\beta > \alpha$. As usual, Greek indices take on the values 0, 1, 2, 3 and Latin indices the values 1, 2, 3. (In (15) we have used a coordinate basis to expand $\widetilde{\mathrm{F}}$, but it is also possible to use a non-coordinate basis $\{\mathbf{d}x^0, \widetilde{\mathbf{a}}^i\}$, where $\{\widetilde{\mathbf{a}}^i\}$ is a non-coordinate basis on the Euclidean space). We rewrite (15) in the following way:

$$\widetilde{\mathrm{F}} = F_{0i} \, \mathbf{d}x^0 \wedge \mathbf{d}x^i + F_{|ij|} \, \mathbf{d}x^i \wedge \mathbf{d}x^j = ( F_{i0}\mathbf{d}x^i ) \wedge \mathbf{d}x^0 + F_{|ij|} \, \mathbf{d}x^i \wedge \mathbf{d}x^j . \tag{16}$$

In (16), $\mathbf{d}x^i$ is to be interpreted as a four-dimensional 1-form field, but $\mathbf{d}x^i$ also denotes a corresponding three-dimensional 1-form field defined on some of the time slices. On these slices $\{\mathbf{d}x^i\}$ forms a basis for 1-form fields and $\{\mathbf{d}x^i \wedge \mathbf{d}x^j\}$ a basis for 2-form fields. Therefore, an arbitrary field $\widetilde{\mathrm{F}}$ can be characterized by a continuous set of three-dimensional 1-form fields, $\{\widetilde{\mathrm{E}}_{x^0}\}$, and a set of three-dimensional 2-form fields $\{\widetilde{\widetilde{\mathrm{H}}}_{x^0}\}$. (The reason for choosing the letters E and H, respectively, is given below.) On slice $m_{x^0}$ the fields satisfy:

$$\widetilde{\mathrm{E}}_{x^0} = E_i(x^0)\mathbf{d}x^i \text{ with } E_i(x^0) = F_{i0}(x^0) \;\; , \widetilde{\widetilde{\mathrm{H}}}_{x^0} = H_{|ij|}(x^0) \, \mathbf{d}x^i \wedge \mathbf{d}x^j \text{ with } H_{ij}(x^0) = F_{ij}(x^0) \;\; . \tag{17}$$

(It can be pointed out here that the components of a field are functions of the points in the manifold on which the field is defined, but that this dependence, mostly, is suppressed in the formulas. Thus, $E_i(x^0)$ denote the components of the field $\widetilde{\mathrm{E}}_{x^0}$ on $m_{x^0}$, but the dependence of the points forming $m_{x^0}$ is suppressed.) In (17) we have used a convention that 2-forms on three-dimensional spaces are denoted by double tilde signs. Then, on three-dimensional spaces, a single tilde means a 1-form, except for the volume-form $\widetilde{\omega}$ (which is a 3-form); double tildes denote a 2-form; a single bar means a vector, except for the 3-vector $\overline{\overline{\omega}}$, and double bars denote a 2-vector. If the coordinate system $\{x^i\}$ is the Cartesian system, the components $E_i(x^0)$ and $H_{ij}(x^0)$ in (17) are obtained from the component matrix in Figure 3A. The components $E_i(x^0)$ are given by the quantities $X, Y, Z$, which - as was noted in section 2 - in traditional theory are the components of a vector describing the electric field. The components $H_{ij}(x^0)$ are given by the quantities $L, M, N$, traditionally the components of a pseudo-vector describing the magnetic field. The unit of both the E- and H-field is Ampere/m. The H-field of this paper times $\mu_0$ (the permeability of free space) corresponds to the B-field of the mks system, and our E-field times $Z_0$ (the impedance of free space) corresponds to the customary electric field in the mks system.

It was shown above that the four-dimensional electromagnetic field represented as the 2-form $\widetilde{\mathrm{F}}$ is sliced into a set of 1-form fields, $\{\widetilde{\mathrm{E}}_{x^0}\}$, and a set of 2-form fields, $\{\widetilde{\widetilde{\mathrm{H}}}_{x^0}\}$, defined on the set of three-dimensional time slices $\{m_{x^0}\}$. To retrieve $\widetilde{\mathrm{F}}$, at a coordinate $x^0$, from the sliced parts, we use the expression

$$\widetilde{\mathrm{F}} = \widetilde{\mathrm{E}}_{x^0} \wedge \mathbf{d}x^0 + \widetilde{\widetilde{\mathrm{H}}}_{x^0} , \tag{18}$$

and interpret the basis fields of the E- and H-field as extended into Minkowski spacetime.

The fields in the sets $\{\widetilde{\mathrm{E}}_{x^0}\}$ and $\{\widetilde{\widetilde{\mathrm{H}}}_{x^0}\}$ are defined on different time slices $m_{x^0}$. But these spaces have identical metric, and we use time-independent coordinate and basis systems. In the expressions for the fields on the slices, $\widetilde{\mathrm{E}}_{x^0} = E_i(x^0)\mathbf{d}x^i$, $\widetilde{\widetilde{\mathrm{H}}}_{x^0} = H_{|ij|}(x^0) \, \mathbf{d}x^i \wedge \mathbf{d}x^j$, we can interpret the 1-form fields $\mathbf{d}x^i$ as 1-form fields on one and the same slice. We then have "mathematically transferred" the fields on the different slices to fields on a single space. Considered in this way, the four-dimensional field $\widetilde{\mathrm{F}}$ has been decomposed into two sequences of three-dimensional fields, $\{\widetilde{\mathrm{E}}_{x^0}\}$ and $\{\widetilde{\widetilde{\mathrm{H}}}_{x^0}\}$, defined on a single Euclidean space.

Seen from an outside perspective, stationary Lorentz observers relative the system $\{m_{x^0}\}$ traverse spacetime along $x^0$-coordinate lines, which are their world lines, and cross the slices with the E- and H-fields. According to the standard view, the observers perceive this as a passage of the time $x^0$ in a single three-dimensional Euclidean space, and where things change with this time. Thus, from the observers inside perspective, the sliced parts of $\widetilde{\mathrm{F}}$ on the different slices are described as two time-varying fields, $\widetilde{\mathrm{E}}_{x^0}$ and $\widetilde{\widetilde{\mathrm{H}}}_{x^0}$, on a single Euclidean space. These two time-dependent fields correspond to just mentioned "mathematically transferred" sequences.

But in a similar way as the four-dimensional electromagnetic field can appear as different field variants, the three-dimensional E- and H-field have four field variants from which one can choose in order to represent the fields. On three-dimensional spaces, 1-forms, 2-forms, vectors and 2-vectors have three independent components. An E-field or a H-field therefore appears as a 1-form field, a 2-form field, a vector field or a 2-vector field. The transformations between the different variants are performed by index rising or index lowering operators and dual operators, see Figure 5. Therefore, from the inside perspective of Lorentz observers, an electromagnetic field is described by two time-varying fields on the ordinary space, the E-field and the H-field, where each of them can be represented by some of the field variants shown in Figure 5.

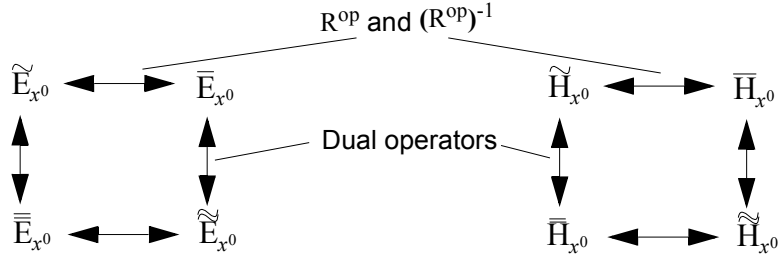

Figure 5. Field variants for E- and H-field.

After having sliced the 2-form field $\widetilde{\mathrm{F}}$, we now show how the other 2-form representation, $\widetilde{\mathrm{F}}^{\iota}$, is sliced. $\widetilde{\mathrm{F}}^{\iota}$ is connected to $\overline{\mathrm{F}}$ with an index raising operation followed by a dual operation:

$$\widetilde{\mathrm{F}}^{\iota} = {}^{*}\overline{\mathrm{F}} = {}^{*}\mathrm{R}^{\mathrm{op}}\overline{\mathrm{F}}^{\iota},\tag{19}$$

where $\overline{\mathrm{F}}$ is a 2-vector (and the star in $\widetilde{\mathrm{F}}^{\iota}$ is a label and not an operator). Comparing with the slicing of $\widetilde{\mathrm{F}}^{\iota}$, it follows from (17) that the components for the sliced 1-form part of $\widetilde{\mathrm{F}}^{\iota}$, on $m_{x^0}$, are equal to $\dot{F}_{i0}(x^0)$. Using the transformation relation (19), and the definitions of the dual and index raising operators in Appendix, yields

$$\dot{F}_{i0}(x^0) = \frac{\Omega_{jki0}}{2} G^{jl}G^{km}F_{lm}(x^0).\tag{20}$$

$\Omega_{jki0}$ are components of the Minkowski volume-form, and $G^{il}$ components of the inverse metric tensor of Minkowski spacetime. The volume-form $\widetilde{\Omega}$ can be written (see Appendix)

$$\widetilde{\Omega} = \mathbf{d}x^0 \wedge \widetilde{\omega},\tag{21}$$

where $\widetilde{\omega}$ is the Euclidean volume-form, $\widetilde{\omega} = \mathbf{d}x \wedge \mathbf{d}y \wedge \mathbf{d}z$, here extended into Minkowski spacetime. The contraction on $\widetilde{\Omega}$ with $\partial/\partial x^0$ (on first slot) satisfies $\widetilde{\Omega}(\partial/\partial x^0, \ , \ , \ ) = \widetilde{\omega}( \ , \ , \ )$. (This follows from the contraction formula (63) in Appendix.) Therefore $\Omega_{0ijk} = \omega_{ijk}$. From (14) we have $G^{jk} = g^{jk}$, and from (17) it follows that $F_{lm}(x^0) = (\widetilde{\mathrm{H}}_{x^0})_{lm}$. Substituting these relations into (20) gives

$$\dot{F}_{i0}(x^0) = -(\widetilde{\mathrm{H}}_{x^0})_i.$$

In a similar way it can be shown that

$$\dot{F}_{ij}(x^0) = (\widetilde{\overset{\approx}{\mathrm{E}}}_{x^0})_{ij}.$$

Thus, $\widetilde{\mathrm{F}}^{\iota}$ is sliced into a set of 1-forms, $\{-\widetilde{\mathrm{H}}_{x^0}\}$, and a set of 2-forms, $\{\widetilde{\overset{\approx}{\mathrm{E}}}_{x^0}\}$, where the fields in the sets are defined on the different time slices $m_{x^0}$. But as was earlier explained, these sets are perceived by stationary Lorentz observers as time-dependent fields on a single Euclidean space. To retrieve $\widetilde{\mathrm{F}}^{\iota}$, at the coordinate $x^0$, from the sliced parts, we use the expression

$$\widetilde{\mathrm{F}}^{\iota} = -\widetilde{\mathrm{H}}_{x^0} \wedge \mathbf{d}x^0 + \widetilde{\overset{\approx}{\mathrm{E}}}_{x^0},\tag{22}$$

and interpret the basis fields of the E- and H-field as extended into Minkowski spacetime.

The remaining field to slice is the current density field J. Because we in the next section will decompose the general laws in (10), where J is represented as the 3-form ${}^{*}\tilde{\mathrm{J}}$, the sliced parts of ${}^{*}\tilde{\mathrm{J}}$ are determined. We first note that a dual operation on a vector is equal to the contraction (on first slot) between the volume-form and the vector, thus

$${}^{*}\tilde{\mathrm{J}} = \widetilde{\Omega}(\tilde{\mathrm{J}}, \ , \ , \ ),\ \text{in shorthand notation,}\ {}^{*}\tilde{\mathrm{J}} = \widetilde{\Omega}(\tilde{\mathrm{J}}).\tag{23}$$

Using the contraction formula (63) in Appendix with $\widetilde{\mathrm{B}} = \mathbf{d}x^0$, $\widetilde{\mathrm{A}} = \widetilde{\omega}$ and $\overline{\mathrm{C}} = \tilde{\mathrm{J}}$, combined with the relations (23) and (21) gives

$${}^{*}\tilde{\mathrm{J}} = \mathbf{d}x^0(\tilde{\mathrm{J}}) \wedge \widetilde{\omega} - \mathbf{d}x^0 \wedge \widetilde{\omega}(\tilde{\mathrm{J}}).\tag{24}$$

Expanding $\bar{J}$ according to:

$$\bar{J} \;=\; c\rho\frac{\partial}{\partial x^0} + \bar{j}\,, \text{ where } \bar{j} = j_i\,\frac{\partial}{\partial x^i}\,,\tag{25}$$

and substituting this into (24), noticing that $\tilde{\omega}(\bar{j})$ is a 2-form, yields

$$^*\,\bar{J} \;=\; -\tilde{\omega}(\dot{j}) \wedge \mathbf{d}x^0 + c\rho\,\tilde{\omega}\,.\tag{26}$$

Thus, the sliced parts of $^*\,\bar{J}$ on $m_{x^0}$ are a 2-form $-\tilde{\omega}(\dot{j}_{x^0})$, and a 3-form $c\rho_{x^0}\,\tilde{\omega}$. Because $\tilde{\omega}(\dot{j}_{x^0}) = {}^*\,\bar{j}_{x^0} = \overset{\approx}{\bar{j}}_{x^0}$, it follows that the sliced parts of $^*\,\bar{J}$ are

$$\text{a 2-form } -\overset{\approx}{\bar{j}}_{x^0} \text{ and a 3-form } c\rho_{x^0}\,\tilde{\omega}\,.\tag{27}$$

## B. Slicing of the general electromagnetic laws

According to relation (10) in section 2, the general laws can be expressed as

$$\mathbf{d}\widetilde{F} = 0,\; \mathbf{d}\widetilde{\mathcal{F}} - {}^*\,\bar{J} = 0.$$

By slicing the 3-forms $\mathbf{d}\widetilde{F}$ and $\mathbf{d}\widetilde{\mathcal{F}} - {}^*\,\bar{J}$ with the same type of procedure as in section 3A, and then putting the sliced parts equal to zero, we obtain a system of four time-dependent relations between three-dimensional fields defined on one Euclidean space. These relations are an inside perspective on the general laws of electromagnetism as perceived by Lorentz observers inside the Minkowski spacetime.

We use the same basis $\{\mathbf{d}x^\alpha\}$ as earlier and expand $\widetilde{F}$ according to:

$$\widetilde{F} \;=\; F_{|\alpha\beta|}\,\mathbf{d}x^\alpha \wedge \mathbf{d}x^\beta.$$

Then (see the definition of exterior derivative in Appendix)

$$\mathbf{d}\widetilde{F} = \frac{\partial F_{|\alpha\beta|}}{\partial x^\gamma}\,\mathbf{d}x^\gamma \wedge \mathbf{d}x^\alpha \wedge \mathbf{d}x^\beta = \frac{\partial F_{|\alpha\beta|}}{\partial x^0}\,\mathbf{d}x^0 \wedge \mathbf{d}x^\alpha \wedge \mathbf{d}x^\beta + \frac{\partial F_{|\alpha\beta|}}{\partial x^i}\,\mathbf{d}x^i \wedge \mathbf{d}x^\alpha \wedge \mathbf{d}x^\beta =$$

$$= \frac{\partial F_{|ij|}}{\partial x^0}\,\mathbf{d}x^i \wedge \mathbf{d}x^j \wedge \mathbf{d}x^0 + \frac{\partial F_{0j}}{\partial x^i}\,\mathbf{d}x^i \wedge \mathbf{d}x^0 \wedge \mathbf{d}x^j + \frac{\partial F_{|jk|}}{\partial x^i}\,\mathbf{d}x^i \wedge \mathbf{d}x^j \wedge \mathbf{d}x^k =$$

$$= \left(\frac{\partial F_{|ij|}}{\partial x^0}\,\mathbf{d}x^i \wedge \mathbf{d}x^j + \frac{\partial F_{j0}}{\partial x^i}\,\mathbf{d}x^i \wedge \mathbf{d}x^j\right) \wedge \mathbf{d}x^0 + \frac{\partial F_{|jk|}}{\partial x^i}\,\mathbf{d}x^i \wedge \mathbf{d}x^j \wedge \mathbf{d}x^k\,.$$

Using relation (17), and noticing that $\mathbf{d}x^i$ is independent of $x^0$, we obtain the sliced parts of $\mathbf{d}\widetilde{F}$ on $m_{x^0}$ as

$$\text{a 2-form } \frac{\partial}{\partial x^0}\widetilde{\widetilde{H}}_{x^0} + \mathbf{d}\widetilde{E}_{x^0} \text{ and a 3-form } \mathbf{d}\widetilde{\widetilde{H}}_{x^0}\,.\tag{28}$$

(Here $\mathbf{d}$ is the exterior derivative acting on a three-dimensional space.) According to the reasoning in the previous section, a Lorentz observer inside spacetime perceives the fields on the different time slices as time-varying fields on a single Euclidean space. Therefore, from the inside perspective, the relation $\mathbf{d}\widetilde{F} = 0$ is described as two time-dependent geometric relations,

$$\frac{\partial}{\partial x^0}\widetilde{\widetilde{H}}_{x^0} + \mathbf{d}\widetilde{E}_{x^0} = 0,\; \mathbf{d}\widetilde{\widetilde{H}}_{x^0} = 0,\tag{29}$$

on one three-dimensional Euclidean space, and where geometric here means coordinate-free with respect to that space.

We now consider the relation $\mathbf{d}\widetilde{\mathcal{F}} - {}^*\,\bar{J} = 0$. Because (as was shown in section 3A) the sliced parts of the fields $\widetilde{F}$ and $\widetilde{\mathcal{F}}$ are the fields $\{\widetilde{E}_{x^0}\}$, $\{\widetilde{\widetilde{H}}_{x^0}\}$, and $\{-\widehat{H}_{x^0}\}$, $\{\widetilde{\widetilde{E}}_{x^0}\}$ respectively, it follows from (28) that the sliced parts of $\mathbf{d}\widetilde{\mathcal{F}}$ on $m_{x^0}$ are

$$\text{a 2-form } \frac{\partial}{\partial x^0}\widetilde{\widetilde{E}}_{x^0} - \mathbf{d}\widehat{H}_{x^0} \text{ and a 3-form } \mathbf{d}\widetilde{\widetilde{E}}_{x^0}\,.\tag{30}$$

By combining these parts with corresponding parts of $^*\,\bar{J}$ in (27), we obtain the inside perspective on the relation $\mathbf{d}\widetilde{\mathcal{F}} - {}^*\,\bar{J} = 0$ as the following two time-dependent geometric relations defined on the Euclidean space:

$$\mathbf{d}\widehat{H}_{x^0} - \frac{\partial}{\partial x^0}\widetilde{\widetilde{E}}_{x^0} = \overset{\approx}{\bar{j}}_{x^0}\,,\; \mathbf{d}\widetilde{\widetilde{E}}_{x^0} = c\rho_{x^0}\,\tilde{\omega}\,. \text{ Together with } \frac{\partial}{\partial x^0}\widetilde{\widetilde{H}}_{x^0} + \mathbf{d}\widetilde{E}_{x^0} = 0\,,\; \mathbf{d}\widetilde{\widetilde{H}}_{x^0} = 0,\tag{31}$$

which was earlier derived, these four relations describe the laws of classical electromagnetism from the inside perspective of a Lorentz observer inside the Minkowski spacetime.

In Maxwell's "A Treatise on Electricity and Magnetism", he uses the concepts "intensity vectors" and "flux vectors" for electrical quantities that he considers are natural to define with reference to line segments and area segments, respectively. He also points out that there are cases where a quantity can be defined with reference to line segments as well as with reference to area segments. In the equation system (31), 1-form fields correspond to intensity vectors and 2-form fields to flux vectors. Here these types of field follow automatically in the derivation of the equation system. Because a 1-form is an operator with one "slot" for a vector argument, and the vector can be considered as a directed line segment, we can say the 1-form is defined from its responses to directed line segments. A 2-form is an operator on two vector arguments, and because the two vectors can form a parallelogram area, see Figure 6A, a 2-form can be said to be defined from its responses to directed area segments.

A typical flux quantity according to Maxwell is the current density, and fully in line with this, the current density in (31) appears as a 2-form field $\overset{\approx}{j}_{x^0}$ (more detailed $\overset{\approx}{j}_{x^0}(\ ,\ )$). If we let $\overline{e}_1\Delta\lambda^1$ and $\overline{e}_2\Delta\lambda^2$ be two small vectors (where $\Delta\lambda^1$ and $\Delta\lambda^2$ are small positive numbers) and operate with $\overset{\approx}{j}_{x^0}$ on these, we obtain the current flux, $\Delta I$, through the parallelogram in Figure 6A. Because $\overset{\approx}{j}_{x^0} = {}^*\overline{j}_{x^0} = \widetilde{\omega}(\overline{j}_{x^0},\ ,\ )$, we have

$$\Delta I = \overset{\approx}{j}_{x^0}(\overline{e}_1\Delta\lambda^1,\ \overline{e}_2\Delta\lambda^2) = \widetilde{\omega}(\overline{j}_{x^0},\overline{e}_1\Delta\lambda^1,\ \overline{e}_2\Delta\lambda^2) = \widetilde{\omega}(\overline{e}_1\Delta\lambda^1,\ \overline{e}_2\Delta\lambda^2,\overline{j}_{x^0}),\tag{32}$$

i.e. equal to the volume (with sign) of the parallelepiped in Figure 6B. Usually this volume is expressed as $\overline{n}\cdot\overline{j}_{x^0}\,\Delta A$, where $\overline{n}$ is the unit normal to the parallelogram formed by the vectors $\overline{e}_1\Delta\lambda^1$, $\overline{e}_2\Delta\lambda^2$, and $\Delta A$ is the area (a positive number) of the parallelogram. It is observed that the definition of $\overline{n}$ gives an ambiguity. We therefore also demand that the vectors $\overline{e}_1\Delta\lambda^1$, $\overline{e}_2\Delta\lambda^2$, $\overline{n}$ form a right-handed system. Then the following relation is valid:

$$\Delta I = \overset{\approx}{j}_{x^0}(\overline{e}_1\Delta\lambda^1,\overline{e}_2\Delta\lambda^2) = \overline{j}_{x^0}\cdot\overline{n}\ \Delta A\ .\tag{33}$$

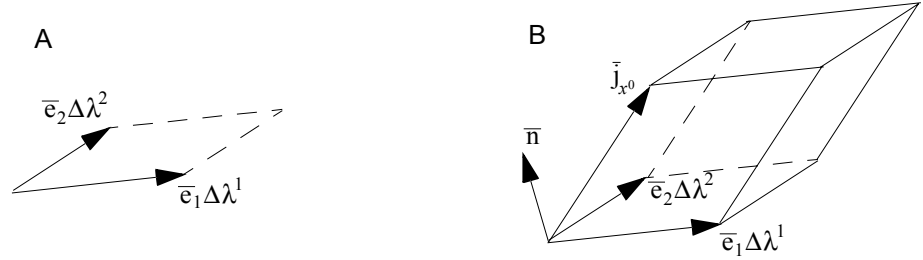

Figure 6. Parallelogram and parallelepiped formed by vectors.

The current flux, $I$, through a (curved) surface S is obtained by summing up contributions from small area segments, defined by small tangential vectors $\overline{e}_1\Delta\lambda^1$ and $\overline{e}_2\Delta\lambda^2$, covering the surface. In the limit where $\Delta\lambda^1$ and $\Delta\lambda^2$ approach zero we have

$$I = \int_S \overset{\approx}{j}_{x^0}(\overline{e}_1,\overline{e}_2)\ d\lambda^1\,d\lambda^2\ .\tag{34}$$

Because the integral expression is valid for arbitrary tangential vector systems covering the surface, the integral of an arbitrary 2-form field, $\overset{\approx}{A}$, over a (two-dimensional) surface is often denoted

$\int_S \overset{\approx}{A}(\ ,\ )$ . In a similar way, the integral of a 1-form field $\widetilde{C}$ over a curve $L$ is written $\int_L \widetilde{C}(\ )$ .

The following generalized Stokes' theorem applies, where $\partial S$ denotes the boundary line of the surface $S$:

$$\int_S \mathbf{d}\widetilde{C}(\ ,\ ) = \int_{\partial S}\widetilde{C}(\ )\ , \text{ or alternatively, } \int_S \mathbf{d}\widetilde{C}(\overline{e}_1,\overline{e}_2)\ d\lambda^1 d\lambda^2 = \int_{\partial S}\widetilde{C}(\overline{e})dl\ .\tag{35}$$

To the formula in (35) must be added a sign rule, see e.g. reference [7], that determines the direction of the tangential vector field $\overline{e}$ on $\partial S$ from the vector basis $\{\overline{e}_1,\overline{e}_2\}$ on $S$. Apart from that, $\overline{e}$ is an arbitrary basis on the boundary line $\partial S$. The formula in (35) can straightforwardly be extended to forms of arbitrary rank. For a 2-form $\overset{\approx}{A}$, we have (where $\partial V$ denotes the boundary surface of the volume $V$)

$$\int_V \mathbf{d}\overset{\approx}{A}(\ ,\ ,\ ) = \int_{\partial V}\overset{\approx}{A}(\ ,\ )\ .\tag{36}$$

If we apply (35) to the first relation in (31), we get the result that the current flux plus the time derivative of the E-field flux through a surface is equal to the line integral of the H-field along the boundary to that surface. Applying (36) to the second relation in (31) gives the result that the E-field flux through a closed surface is equal to the charge inside the surface multiplied by the velocity of light.

Of course there are a lot more to describe about the equation system (31) and about electromagnetism in differential-form notation, than what is done here. See the reference list which gives some examples of references dealing with electromagnetism in differential-form notation.

## C. Transforming of equation systems. A pseudo-free vector analysis

The equation system derived in the preceding section contains fields represented as 1-forms, 2-forms and 3-forms. The system is called the A-system and is given by

$$\text{A:} \ \mathbf{d}\widetilde{E}_{x^0} + \frac{\partial}{\partial x^0}\widetilde{\widetilde{H}}_{x^0} = 0 \ , \ \mathbf{d}\widetilde{\widetilde{H}}_{x^0} = 0 \ , \ \mathbf{d}\widetilde{H}_{x^0} - \frac{\partial}{\partial x^0}\widetilde{\widetilde{E}}_{x^0} = \widetilde{\widetilde{J}}_{x^0} \ , \ \mathbf{d}\widetilde{\widetilde{E}}_{x^0} = c\rho_{x^0}\widetilde{\omega} \ . \tag{37}$$

But we can choose from a set of four field variants in order to represent a field, see Figure 5 which shows the different variants for the E- and H-field. Also the current density field has the same set of field variants, while the charge density can appear as a 3-form, a 3-vector or a scalar (which is the same as a 0-form and 0-vector). Because a 3-form on a three-dimensional space has one independent component, the charge density field represented as a 3-form can be written $\rho_{x^0}\widetilde{\omega}$, where $\rho_{x^0}$ is the scalar charge density field and $\widetilde{\omega}$ is the volume-form. As follows from relation (72) in Appendix, the dual of $\rho_{x^0}\widetilde{\omega}$ is the scalar field $\rho_{x^0}$, i.e. $^*(\rho_{x^0}\widetilde{\omega}) = \rho_{x^0}$. We now operate with a dual operator on both sides of the relations in (37), and use the connections

$$^*\widetilde{\widetilde{H}}_{x^0} = \overline{H}_{x^0} \ , \ ^*\overline{H}_{x^0} = \widetilde{\widetilde{H}}_{x^0} \ , \ ^*\widetilde{\widetilde{E}}_{x^0} = \overline{E}_{x^0} \ , \ ^*\overline{E}_{x^0} = \widetilde{\widetilde{E}}_{x^0} \ , \ ^*\widetilde{\widetilde{J}}_{x^0} = \overline{J}_{x^0} \ , \ ^*(\rho_{x^0}\widetilde{\omega}) = \rho_{x^0} \ , \ ^*\frac{\partial}{\partial x^0} = \frac{\partial}{\partial x^0}{}^* \ .$$

(The commutation relation applies because the metric tensor and the volume-form, which define the dual operators, are time-independent.) The result is an alternative equation system

$$\text{B:} \ ^*\mathbf{d}\widetilde{E}_{x^0} + \frac{\partial}{\partial x^0}\overline{H}_{x^0} = 0 \ , \ ^*\mathbf{d}\ ^*\overline{H}_{x^0} = 0 \ , \ ^*\mathbf{d}\widetilde{H}_{x^0} - \frac{\partial}{\partial x^0}\overline{E}_{x^0} = \overline{J}_{x^0} \ , \ ^*\mathbf{d}\ ^*\overline{E}_{x^0} = c\rho_{x^0} \ . \tag{38}$$

The fields in this system are represented as 1-forms, vectors and scalars. The 1-form fields can be transformed into vector fields with an index raising operator, $R^{op}$. We have $\widetilde{H}_{x^0} = (R^{op})^{-1}\ \overline{H}_{x^0}$, $\widetilde{E}_{x^0} = (R^{op})^{-1}\ \overline{E}_{x^0}$, which can be expressed with the aid of the metric tensor $\mathbf{g}$ as $\widetilde{H}_{x^0} = \mathbf{g}(\ ,\overline{H}_{x^0})$, $\widetilde{E}_{x^0} = \mathbf{g}(\ ,\overline{E}_{x^0})$. Substituting these relations into the equation system (38) yields a system with vector and scalar fields, and we have

$$\text{C:} \ \nabla \vee \overline{E}_{x^0} + \frac{\partial}{\partial x^0}\overline{H}_{x^0} = 0 \ , \ \nabla \cdot \overline{H}_{x^0} = 0 \ , \ \nabla \vee \overline{H}_{x^0} - \frac{\partial}{\partial x^0}\overline{E}_{x^0} = \overline{J}_{x^0} \ , \ \nabla \cdot \overline{E}_{x^0} = c\rho_{x^0} \ . \tag{39}$$

The operators $\nabla \vee$ and $\nabla \cdot$ are composed of basic operators according to:

$$\nabla \cdot = \ ^*\mathbf{d}\ ^* \ , \ \nabla \vee = \ ^*\mathbf{d}(R^{op})^{-1} \ , \text{ where here } (R^{op})^{-1} \text{ is defined by } (R^{op})^{-1}\ \overline{A} = \mathbf{g}(\ ,\overline{A}) = \widetilde{A} \ . \tag{40}$$

See Figure 1 which gives an overview of the different equation systems that have been derived. The operator $\nabla \cdot$ operates on vector fields, resulting in scalar fields, and the operator $\nabla \vee$ operates on vector fields, resulting in other vector fields. Therefore the E- and H-field of the C-system are vector fields of the same sort, which is quite natural because there is only one sort of vector field in exterior calculus. The reason for pointing this out is that in traditional Maxwell's equations, with div and curl operators, the electric field and the magnetic field are described by vectors of different sorts. The electric field vector is a "true" vector whereas the magnetic field vector is a "pseudo" vector. One also has to use true scalar fields and pseudo-scalar fields. The need for introducing these pseudo-concepts depends on an insufficient definition of the curl operator. If the curl operator operates on a vector field the result is a pseudo-vector field. Such fields do not transform in the same way as true vector fields for basis transformations containing a change in the handedness of the basis system. In section E, on mirror transformations and mirror symmetry, these questions are touched on somewhat further.

An important theorem in traditional electromagnetic theory is Stokes' theorem, $\iint \text{curl}(\overline{A}) \cdot \overline{n} \ dS = \oint \overline{A} \cdot d\overline{s}$. We now show how a corresponding theorem with no pseudo-quantities (and thus adapted to the C-system) can be derived. Using (40) and (33) gives

$$(\nabla \vee \overline{A}) \cdot \overline{n} \ \Delta S = (\ ^*\mathbf{d}(R^{op})^{-1}\ \overline{A}) \cdot \overline{n} \ \Delta S = (\ ^*\mathbf{d}\widetilde{A}) \cdot \overline{n} \ \Delta S = (\ ^*\ ^*\mathbf{d}\widetilde{A})(\overline{e}_1\Delta\lambda^1, \overline{e}_2\Delta\lambda^2) \ ,$$

where $\overline{e}_1\Delta\lambda^1$, $\overline{e}_2\Delta\lambda^2$, $\overline{n}$ must form a right-handed system. Apart from that, $\{\overline{e}_1, \overline{e}_2\}$ is an arbitrary tangential vector system on $S$. From the generalized Stokes' theorem in (35), the last relation in (40) and because $^*\ ^*$ is equal to the identity operator (see relation (71) in Appendix), it then follows that

$$\int_S (\nabla \vee \overline{A}) \cdot \overline{n} \ dS = \int_{\partial S} \widetilde{A}(\overline{e}) \ dl = \int_{\partial S} \mathbf{g}(\overline{e}, \overline{A}) \ dl = \int_{\partial S} \overline{A} \cdot \overline{e} \ dl \ . \tag{41}$$

$\overline{e}$ is an arbitrary tangential system on the boundary line $\partial S$, except that the direction of $\overline{e}$ must satisfy the sign rule of the generalized Stokes' theorem. This rule together with the right-handed system above give a connection between up or down for the unit normal $\bar{n}$ on the surface $S$, and the integration direction on the boundary line $\partial S$. With the aid of the generalized Stokes' theorem in (36) a Gauss's theorem can, in a similar way, be derived:

$$\int_V \nabla \cdot \bar{A} \, dv = \int_{\partial V} \bar{A} \cdot \bar{n} \, dS \,. \tag{42}$$

$(\nabla \cdot)(\nabla \vee) = 0$ is another example of a theorem adapted to the C-system. From (40) we have $(\nabla \cdot)(\nabla \vee) = {}^*\mathbf{d} \, {}^* \, {}^*\mathbf{d}(R^{op})^{-1} = {}^*\mathbf{dd}(R^{op})^{-1} = 0$ (see Appendix for definitions of exterior derivative and dual operators). This corresponds to the traditional theorem that a curl operation followed by a div operation is equal to zero.

A commonly used operation in traditional vector theory is the cross product between vectors or pseudo-vectors. If the cross product operates on two vectors the result is a pseudo-vector. If it operates on a vector and a pseudo-vector the result is a vector. We now define a corresponding operation between two vectors, denoted $\bar{A} \vee \bar{B}$, with no pseudo-quantities involved (and thus adapted to the C-system):

$$\bar{A} \vee \bar{B} = {}^* (\widetilde{A} \wedge \widetilde{B}) = {}^* (\mathbf{g}(\ ,\bar{A}) \wedge \mathbf{g}(\ ,\bar{B})) \,. \tag{43}$$

As a further example of a theorem adapted to the C-system, we show how the correspondence to the traditional theorem $\mathrm{div}(\bar{A} \times \bar{B}) = \bar{B} \cdot \mathrm{curl}(\bar{A}) - \bar{A} \cdot \mathrm{curl}(\bar{B})$ is derived. In addition to earlier given relations and definitions, we also use that $\widetilde{C} \wedge \widetilde{D} = \widetilde{C} \wedge \widetilde{D} = \bar{C} \cdot \bar{D} \, \widetilde{\omega}$ on the three-dimensional Euclidean space. We then have

$$\nabla \cdot (\bar{A} \vee \bar{B}) = {}^*\mathbf{d} \, {}^* ({}^* (\widetilde{A} \wedge \widetilde{B})) = {}^* (\mathbf{d}\widetilde{A} \wedge \widetilde{B}) - {}^* (\widetilde{A} \wedge \mathbf{d}\widetilde{B}) = {}^* (({}^*\mathbf{d}\widetilde{A}) \cdot \bar{B} \, \widetilde{\omega}) -$$
$$- {}^* (({}^*\mathbf{d}\widetilde{B}) \cdot \bar{A} \, \widetilde{\omega}) = ({}^*\mathbf{d}\widetilde{A}) \cdot \bar{B} - ({}^*\mathbf{d}\widetilde{B}) \cdot \bar{A} = \bar{B} \cdot \nabla \vee \bar{A} - \bar{A} \cdot \nabla \vee \bar{B} \,. \tag{44}$$

With the aid of (44), (42) and the C-system in (39) a Poynting relation for the C-system can be obtained.

Hence, a set of theorems can be derived using the operator definitions in (40) and (43). These theorems form a vector analysis with the operators $\nabla \vee$, $\nabla \cdot$, $\nabla$, $\vee$, and the scalar product, that corresponds to the traditional vector analysis with div, curl, grad, cross product and scalar product, but is pseudo-free and adapted to the C-system. (The operator $\nabla$ above is equal to $R^{op}\mathbf{d}$).

The C-system is derived from general four-dimensional electromagnetic laws, it has no pseudo-quantities, and it is a part of a unified theory of classical electromagnetism, schematically described by Figure 1. Therefore, it is a more basic (and rigorous) system compared to the traditional Maxwell's equations with div and curl operators.

### D. Equation systems with magnetic source fields

In the presence of magnetic sources, $\mathbf{J}^M$, the general laws satisfy $\mathbf{d}\widetilde{F} = {}^* \widetilde{J}^M$, $\mathbf{d}\widetilde{F} = {}^* \widetilde{J}$. It follows from a comparison with (27) that the sliced parts of ${}^* \widetilde{J}^M$ on $m_{x^0}$ are

$$\text{a 2-form } -\widetilde{j}_{x^0}^M \text{ and a 3-form } c\rho_{x^0}^M \, \widetilde{\omega} \,. \tag{45}$$

By using the sliced parts of $\mathbf{d}\widetilde{F}$ in relation (28), combined with corresponding parts in (45), we obtain an inside perspective on the relation $\mathbf{d}\widetilde{F} = {}^* \widetilde{J}^M$ as the two relations:

$$\mathbf{d}\widetilde{E}_{x^0} + \frac{\partial}{\partial x^0} \widetilde{H}_{x^0} = -\widetilde{j}_{x^0}^M \,, \ \mathbf{d}\widetilde{H}_{x^0} = c\rho_{x^0}^M \, \widetilde{\omega} \,. \tag{46}$$

Replacing the first two relations of the A-system in (37) with (46) gives an equation system where both electric and magnetic source fields are included. Transforming (46) according to the transformations in previous section yields

$$\nabla \vee \bar{E}_{x^0} + \frac{\partial}{\partial x^0} \bar{H}_{x^0} = -\bar{j}_{x^0}^M \,, \ \nabla \cdot \bar{H}_{x^0} = c\rho_{x^0}^M \,. \tag{47}$$

Replacing the first two relations of the C-system in (39) with (47) gives a system with both electric and magnetic source fields.

### E. Mirror transformations and mirror symmetry

In this section we analyse some properties of the electromagnetic field and its laws with respect to mirror transformations. The analysis will include four-dimensional fields and laws on the Minkowski spacetime, but we start with the inside perspective, with fields and laws defined on the three-dimensional Euclidean space.

A mirror operator, denoted $P^{op}$, is defined in relation to a plan. Here we assume that the plane is the yz-plane in a Cartesian coordinate system, and that $x^1 = x$, $x^2 = y$, $x^3 = z$. We use the vector basis $\{\partial/\partial x^i\}$, the 1-form basis $\{\mathbf{d}x^i\}$ and the 2-form basis $\mathbf{d}x^i \wedge \mathbf{d}x^j$ to expand a vector field $\bar{A}$, a 1-form field $\widetilde{B}$ and a 2-form field $\widetilde{C}$, according to:

$$\bar{A} = A^i \frac{\partial}{\partial x^i} \,, \ \widetilde{B} = B_i \, \mathbf{d}x^i \,, \ \widetilde{C} = C_{[ij]} \, \mathbf{d}x^i \wedge \mathbf{d}x^j \,. \tag{48}$$

We now define the mirror operator as an operator which transforms fields into mirror fields according to the relations below, where the index m denotes a mirror field.

For a scalar field $\Phi$, its mirror field $\Phi_m$ satisfies: $\Phi_m = P^{op}\Phi$, where $\Phi_m(x^1, x^2, x^3) = \Phi(-x^1, x^2, x^3)$.

For a vector field $\overline{A}$, $\overline{A}_m = P^{op}\overline{A}$, where the components satisfy:

$$(A_m)^i(x^1, x^2, x^3) = -A^i(-x^1, x^2, x^3) \quad \text{if } i = 1, \text{ otherwise} \quad (A_m)^i(x^1, x^2, x^3) = A^i(-x^1, x^2, x^3) \ .$$

For a 1-form field $\tilde{B}$, $\tilde{B}_m = P^{op}\tilde{B}$, where the components satisfy:

$$(B_m)_i(x^1, x^2, x^3) = -B_i(-x^1, x^2, x^3) \text{ if } i = 1, \text{ otherwise } (B_m)_i(x^1, x^2, x^3) = B_i(-x^1, x^2, x^3) \ .$$

For a 2-form field $\overset{\approx}{C}$, $\overset{\approx}{C}_m = P^{op}\overset{\approx}{C}$, where the components satisfy:

$$(C_m)_{ij}(x^1, x^2, x^3) = -C_{ij}(-x^1, x^2, x^3) \text{ if } i \text{ or } j \text{ is 1, otherwise} \quad (C_m)_{ij}(x^1, x^2, x^3) = C_{ij}(-x^1, x^2, x^3) \ .$$

The relations for other p-forms and p-vectors are clear from the rules above. The definitions are also, essentially, valid for fields on the Minkowski spacetime. We just have to add a time coordinate and change the Latin indices into Greek indices.

If a field and its mirror field are equal, i.e. $K = P^{op}K$, where K denotes an arbitrary type of field, then the field is mirror symmetric with respect to the mirror plane. Such a field is an eigenfield of the mirror operator, with eigenvalue +1, and is said to have parity +1. If we have $P^{op}K = -K$, the field has parity -1 and is an anti-symmetric field.

With the aid of the definitions above for the mirror operator, the definitions in Appendix for the exterior derivative, the dual operators and the index raising operator, the following commutation relations can be shown to apply:

$$\mathbf{d}P^{op} = P^{op}\mathbf{d} \ , \quad {}^*P^{op} = -P^{op}\, {}^* \ , \quad R^{op}P^{op} = P^{op}\, R^{op} \ , \quad \frac{\partial}{\partial x^0}P^{op} = P^{op}\, \frac{\partial}{\partial x^0} \ . \tag{49}$$

(The last relation applies because the mirror operator is time-independent.) The relations are valid both on the Euclidean space and the Minkowski spacetime.

We now assume an E-field represented as the 1-form field $\tilde{E}_{x^0}$ and that $\tilde{E}_{x^0}$ is a symmetric field, i.e. $P^{op}\tilde{E}_{x^0} = \tilde{E}_{x^0}$. Because $\overline{E}_{x^0} = R^{op}\tilde{E}_{x^0}$, we have from the third commutation relation in (49) that also $\overline{E}_{x^0}$ is a symmetric field. But if we represent the E-field as the 2-form $\overset{\approx}{E}_{x^0} = {}^*\overline{E}_{x^0}$, it follows from relation two in (49) that $\overset{\approx}{E}_{x^0}$ is an antisymmetric field. Because dual operators anticommute with the mirror operator, dual operators change the parity. The rest of the operators in (49) preserve the parity.

From the outside perspective, the electromagnetic field is represented either as the 2-form $\tilde{F}$ or the 2-form $\overset{\approx}{F}$. (The 2-vector fields that also can represent the electromagnetic field are not used in this paper.) The 2-forms are connected to each other by the relation

$$\overset{\approx}{F} = {}^*R^{op}\, \tilde{F} \ . \tag{50}$$

Assume $\tilde{F}$ is a mirror symmetric field, i.e. its parity is +1. Because $R^{op}$ in (50) preserves the parity and the dual operator $^*$ changes it, $\overset{\approx}{F}$ has parity -1 and thus is an antisymmetric field. We now define a mirror symmetric electromagnetic field as a field where $\tilde{F}$ is mirror symmetric. From the general laws (where we have included magnetic sources), $\mathbf{d}\tilde{F} = {}^*\tilde{J}^M$, $\mathbf{d}\overset{\approx}{F} = {}^*\tilde{J}$, it follows that the sources to such a field are a mirror symmetric vector field $\tilde{J}$, representing the electric current density, and an antisymmetric vector field $\tilde{J}^M$, representing the magnetic current density. (This is because $\mathbf{d}$ preserves the parity and $^*$ changes it.)

Using relation (18), $\tilde{F} = \tilde{E}_{x^0} \wedge \mathbf{d}x^0 + \overset{\approx}{H}_{x^0}$, we find that the sliced parts $\tilde{E}_{x^0}$ and $\overset{\approx}{H}_{x^0}$ are mirror symmetric for a mirror symmetric electromagnetic field. (The wedge product $\tilde{E}_{x^0} \wedge \mathbf{d}x^0$ has the same parity as $\tilde{E}_{x^0}$.) If we choose to represent the E- and H-field as vector fields, then the inside perspective on a mirror symmetric electromagnetic field is a symmetric $\overline{E}_{x^0}$ field and an antisymmetric $\overline{H}_{x^0}$ field (because $\overline{E}_{x^0} = R^{op}\tilde{E}_{x^0}$, $\overline{H}_{x^0} = {}^*\overset{\approx}{H}_{x^0}$ and a dual operator changes the parity and an index raising operator preserves it).

We now consider a general electric circuit, exemplified by Figure 7A, and let all fields of the circuit be represented by vector fields and scalar fields. We assume there are no magnetic source fields. Using the C-system in (39) to describe the relations between the field quantities of the circuit in Figure 7A yields (where we have dropped the time index on the fields):

$$\nabla \vee \overline{E}^A + \frac{\partial}{\partial x^0}\overline{H}^A = 0, \ \nabla \cdot \overline{H}^A = 0 \ , \ \nabla \vee \overline{H}^A - \frac{\partial}{\partial x^0}\overline{E}^A = \overline{j}^A, \ \nabla \cdot \overline{E}^A = c\rho^A. \tag{51}$$

Because $\nabla \cdot = {}^*\mathbf{d}\,{}^*$ and $\nabla \vee = {}^*\mathbf{d}(R^{op})^{-1}$, the commutation relations in (49) give the following commutation relations for the operators $\nabla \cdot$ and $\nabla \vee$:

$$\nabla \cdot P^{op} = P^{op}\, \nabla \cdot \ , \quad \nabla \vee P^{op} = -P^{op}\, \nabla \vee. \tag{52}$$

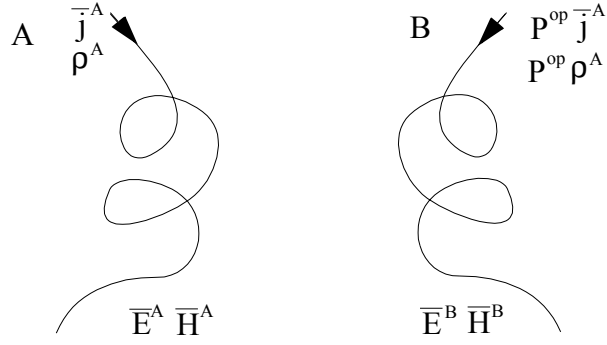

Figure 7. An electric circuit and its mirror circuit.

We now operate with $-P^{op}$ on the first two relations in (51) and with $P^{op}$ on the last two, and use the commutation relations in (52) and number four in (49). This gives

$$\nabla \vee (P^{op}\overline{E}^A) + \frac{\partial}{\partial x^0}(-P^{op}\overline{H}^A) = 0, \ \nabla \cdot (-P^{op}\overline{H}^A) = 0, \tag{53}$$

$$\nabla \vee (-P^{op}\overline{H}^A) - \frac{\partial}{\partial x^0}(P^{op}\overline{E}^A) = P^{op}\overline{j}^A, \ \nabla \cdot (P^{op}\overline{E}^A) = cP^{op}\rho^A. \tag{54}$$

Thus, the electric and magnetic fields, $\overline{E}^B$, $\overline{H}^B$, of the mirror circuit in Figure 7B with the mirror sources $P^{op}\overline{j}^A$ and $P^{op}\rho^A$ satisfy

$$\overline{E}^B = P^{op}\overline{E}^A, \ \overline{H}^B = -P^{op}\overline{H}^A. \tag{55}$$

The explanation to the result in (55) is not that the magnetic field transforms in a different way compared to the electric field for mirror transformations. Both $\overline{E}$ and $\overline{H}$ are vector fields of the same kind, and transform in exactly the same way for all kinds of transformations, including mirror transformations. As follows from the derivation, the explanation to the result in (55) is that the operator $\nabla \vee$ anticommute with the mirror operator while $\partial / \partial x^0$ and $\nabla \cdot$ commute with it.

## 4. SUMMARY

This paper has dealt with a theory of classical electromagnetism built on relativity principles, Maxwell's original equations and the mathematics of exterior calculus. It is a theory that encompasses both the outside and inside perspectives on electromagnetism.

The outside perspective gives a complete overview of electromagnetism for all kinds of spacetimes. From this perspective, the laws are geometric (coordinate-free) relations on a general four-dimensional spacetime. The laws are obtained by applying the principle of general covariance and exterior calculus on Maxwell's original equations. In section 2, the outside perspective was briefly described.

In section 3, which is the central part of the paper, the perspective was narrowed into the perspective of non-accelerating observers inside the ordinary flat spacetime. Based on the general laws, exterior calculus and a slicing of the Minkowski spacetime, we derived fields and equation systems that describe electromagnetism from the perspective of Lorentz observers, inside the Minkowski spacetime. In the primary equation system, A, all fields in the relations are time-dependent form fields. For example, the current density field is represented as a 2-form field $\overset{\approx}{j}_{x^0}$. If it operates on two small vectors, the current flux through the parallelogram formed by the vectors is obtained. Using the operator notion of 2-forms and 1-forms the connections to Maxwell's "flux vectors" and "intensity vectors" were explained.

Since a field of exterior calculus (on metric spaces) is associated to four field variants, the equation system A can be transformed into other systems, where the fields are represented in alternative ways. In the C-system, the fields appear as vector and scalar fields. But there is no need of introducing the concepts pseudo-vector and pseudo-scalar as in the traditional Maxwell's equations with div and curl operators. The C-system is derived from general laws and is a part of a unified theory of electromagnetism encompassing both the outside and inside perspectives. Because of this, it is a more basic (and rigorous) system than the traditional Maxwell's equations. The curl and div operators in the traditional equations correspond to the derived operators $\nabla \vee$ and $\nabla \cdot$ in the C-system. $\nabla \vee$ and $\nabla \cdot$ are composed of the basic operators, exterior derivative, dual operators and an index raising operator. We also introduced an operator $\vee$, corresponding to the traditional cross product operator. Using the definitions of the operators $\nabla \vee$, $\nabla \cdot$, $\vee$, and the scalar product, we derived some relations of a pseudo-free vector analysis adapted to the C-system. For example, a Stokes' theorem containing the $\nabla \vee$ operator and vector fields, and no pseudo fields, was derived.

The paper was concluded with a section on mirror transformations and mirror symmetry. The analysis was built on a set of commutation relations between the mirror operator and exterior calculus operators, including the index raising operator. Using these relations and a definition of a mirror symmetric electromagnetic field, we derived some relations describing mirror properties of different field quantities.

## 6. APPENDIX. Some exterior calculus concepts, relations and notations

This section gives a summary description of exterior calculus concepts, relations and notations that are used in the paper. References have been [7] and [8] in the reference list. Many of the concepts described are not specific for exterior calculus but belong to the wider theory of differential geometry. Some such concepts are spaces, coordinates, vectors and 1-forms. We also describe the metric concept, which is a pure differential geometric concept and does not belong to exterior calculus.

*Manifolds, spaces and coordinates*. Apart from this section, the manifolds of interest in the paper either represent the ordinary three-dimensional Euclidean space or four-dimensional spacetime, but here we consider manifolds with arbitrary dimensions. Without using a very strict mathematical definition, we describe a manifold as a collection of points (objects) where each point, $P$, can be mapped to a real n-tuple $(x^1(P) ,  , ....... , x^n(P) )$ in a one-to-one manner. $n$ is the dimension of the manifold, and the numbers in the n-tuple are the coordinates of $P$ in the coordinate system which defines the mapping between points and n-tuples. (In general it is not possible to cover the whole manifold with one coordinate system and also fulfil the criteria of a one-to-one mapping.) Most often the dependence of $P$ is dropped and the coordinates are denoted $x^i$, where $i = 1, 2, . . . . . . n$. In four-dimensional spacetime we use the standard convention that coordinates have Greek indices which run from 0 to 3, and where index 0 denotes the time coordinate. But in this Appendix (with two exceptions) we use Latin indices ranging from 1 to $n$.

If the manifold satisfies a "smoothness" criteria we have a differentiable manifold, which according to reference [8] is the mathematically precise substitute for the concept space. All the manifolds considered in this paper are such manifolds, and therefore we will use manifold, differentiable manifold and space as synonymous concepts. Furthermore, all the spaces (manifolds) considered in the paper have a metric tensor defined on them. Differentiable manifolds with a metric tensor are called pseudo-Riemannian manifolds or, equivalently, metric spaces.

*Vectors and vector fields*. We assume the curve $\{ x^i = p^i(\lambda) , i = 1, 2, . . . . . . n \}$ passes through the point $P$ and let $f$ be a function that is defined on the manifold. Along the curve we then have a function $h(\lambda) = f(x^1, ......., x^n)$. By differentiating and using the chain rule we obtain

$$\frac{\mathrm{d}h}{\mathrm{d}\lambda} = \sum_i \frac{\mathrm{d}p^i}{\mathrm{d}\lambda} \frac{\partial f}{\partial x^i}.$$

Because this is true for an arbitrary (differentiable) function $f$, we have

$$\frac{\mathrm{d}}{\mathrm{d}\lambda} = \sum_i \frac{dp^i}{d\lambda} \frac{\partial}{\partial x^i},$$

i.e. any directional derivative can be written as a linear combination of directional derivatives along the coordinate axes. Thus, the set $\{\partial/\partial x^i\}$ is a basis for the space of directional derivatives at $P$. This space is called the tangent space at $P$, it is denoted $T_P$, and (in the modern view) the directional derivatives of $T_P$ are vectors at $P$. (Vectors do not belong to the manifold but to local spaces $T_P$.) However, we can use the traditional picture of a vector as an "arrow", with components $dp^i/d\lambda$, directed tangential to the curve $\{x^i = p^i(\lambda)\}$. The term vector field refers to a rule for defining a vector at each point of a manifold, and it can be shown that an arbitrary vector field is associated to a set of manifold filling curves, called the congruence of the field. Thus, the pictorial representation of a vector field is a set of "arrows" fixed on congruence curves and tangential to the curves.

A tangent space $T_P$ has the same dimension, $n$, as the manifold, and any set of $n$ linearly independent vectors in $T_P$ is a basis for that space. By choosing a basis in all the tangent spaces we get a basis for vector fields on the manifold. The set $\{\partial/\partial x^i\}$ is a coordinate vector basis, but other non-coordinate basis, $\{\bar{e}_i\}$, can be used. (The bar indicates a vector quantity.) An arbitrary vector field $\bar{A}$ can then be written:

$$\bar{A} = \sum_i A^i \frac{\partial}{\partial x^i} = \sum_i A^{i'} \bar{e}_i, \tag{56}$$

where $A^i$ denote components of $\bar{A}$ on the coordinate vector basis $\{\partial/\partial x^i\}$ and $A^{i'}$ denote components on the basis $\{\bar{e}_i\}$.

*1-forms and 1-form fields*. A 1-form is defined as a linear real-valued operator acting on vectors. Thus, a 1-form $\tilde{B}$ at a point $P$ (where the tilde is used to indicate form quantities) associates with a vector $\bar{A}$, belonging to $T_P$, a real number $\tilde{B}(\bar{A})$. It can be shown that 1-forms at $P$ satisfy the axioms of a vector space. It is called the dual space to $T_P$ and is denoted $T_P^*$. A 1-form field is a rule which defines a 1-form at each point of the manifold.

*The gradient* of a function $f$ is a 1-form field $\mathbf{d}f$, defined by its action on vector fields according to the relation $\mathbf{d}f(d/d\lambda) = df/d\lambda$.

*Dual basis*. In the vector space of 1-forms, $T_P^*$, any $n$ linearly independent 1-forms define a basis. But if a vector basis $\{\bar{e}_i\}$ has been chosen for $T_P$, a "preferred" basis for $T_P^*$, called the dual basis to $\{\bar{e}_i\}$, can be defined. If we denote this basis $\{\tilde{a}^i\}$, we have

$$\tilde{a}^i(\bar{e}_j) = \delta_j^i,$$

where the Kronecker symbol $\delta_j^i$ is 1 if $i = j$ and 0 for other cases. If $\{\bar{e}_i\}$ is a vector basis field then $\{\tilde{a}^i\}$ is the dual basis field to $\{\bar{e}_i\}$. From the gradient definition it follows that the dual basis field to the coordinate vector basis $\{\partial/\partial x^i\}$ is $\{\mathbf{d}x^i\}$. Thus, an arbitrary 1-form field $\tilde{B}$ can be expressed as

$$\tilde{B} = \sum_i B_i \mathbf{d}x^i. \tag{57}$$

*Exterior calculus* is a theory for studying completely antisymmetric tensors of type $(0/p)$, called p-forms, and completely antisymmetric tensors of type $(p/0)$, called p-vectors. A tilde over a letter indicates a p-form, e.g. $\tilde{A}$, or more detailed $\tilde{A}(\ ,\ ,\ ......,\ )$ with $p$ "slots" for the vector arguments ($p$ is called the rank of the tensor). We thus consider tensors as linear operators which operate on vectors and 1-forms, resulting in real numbers. A bar over a letter indicates a p-vector, e.g. $\bar{B}$, or more detailed $\bar{B}(\ ,\ ,\ ......,\ )$ with $p$ slots for the 1-form arguments. The antisymmetry means that the value of the tensor, when operating on the arguments, changes sign on interchange of any two arguments. A 1-vector is the same as a vector, and 0-vectors and 0-forms the same as scalars. A p-form field or p-vector field is a rule for determining a p-form or p-vector at each point of the manifold (space) on which the tensors are defined. (But often we will use the designation p-form, p-vector and tensor instead of p-form field, p-vector field and tensor field.) The different types of tensors do not belong to the manifold, but have their own local spaces; e.g. the tangent spaces $T_P$ of vector fields.

*Index and summation conventions*. Components of $(p/0)$-tensors have the indices written as superscript, and components of $(0/p)$-tensors have subscripted indices. Members of vector basis are labelled with subscripts, and those of 1-form basis with superscript. When an expression contains a repeated index, once as subscript and once as superscript, a summation over the index is understood. By using these conventions, the fields in (56) and (57) are expressed as:

$$\bar{A} = A^i \frac{\partial}{\partial x^i} = A^{i'} \bar{e}_i, \ \tilde{B} = B_i \mathbf{d}x^i. \tag{58}$$

*Metric tensors. Inner product (scalar product)*. The manifolds of interest in this paper are metric manifolds (metric spaces), i.e. manifolds which have a metric tensor defined on them, describing the metric properties of the manifold. This tensor, which we denote $\mathbf{G}$, or more detailed $\mathbf{G}(\quad,\quad)$, for spacetime manifolds and $\mathbf{g}$ (alternatively $\mathbf{g}(\quad,\quad)$) for other cases, is a symmetric (0/2)-tensor. The symmetry means that $\mathbf{g}(\overline{A},\overline{B}) = \mathbf{g}(\overline{B},\overline{A})$ for arbitrary vectors $\overline{A}$ and $\overline{B}$. $\mathbf{g}(\overline{A},\overline{B})$ is also denoted $\overline{A}\cdot\overline{B}$ and is defined as the inner product of $\overline{A}$ and $\overline{B}$. The components of $\mathbf{g}$ on a vector basis $\{\overline{e}_i\}$ are $g_{ij} = \mathbf{g}(\overline{e}_i,\overline{e}_j)$. The corresponding matrix has to be invertible, where the components of the inverse matrix, $g^{ij}$, can be shown to be components of a symmetric (2/0)-tensor which we denote $\mathbf{g}^{-1}$. With these two tensors, vectors can be mapped into 1-forms and vice versa in a one-to-one manner. For a given vector $\overline{A}$, $\mathbf{g}(\quad,\overline{A})$ is a 1-form which we denote $\tilde{A}$. Thus, $\tilde{A} = \mathbf{g}(\quad,\overline{A})$ and $\overline{A} = \mathbf{g}^{-1}(\quad,\tilde{A})$. Expressed in components we have $A_i = g_{ij}A^j$, $A^i = g^{ij}A_j$, where the index and summation conventions have been used.

*Index raising and index lowering operators.* The maps defined above, called index raising and index lowering operations, between vectors (1-vectors) and 1-forms can be extended to maps between p-vectors and p-forms (it can be shown that the antisymmetry is preserved by these maps). For example, if $\tilde{A}$ is a 2-form with the components $A_{ij} = \tilde{A}(\overline{e}_i,\overline{e}_j)$ on a vector basis $\{\overline{e}_i\}$, then the associated 2-vector $\overline{A}$ has the components $A^{ij} = g^{il}g^{jk}A_{lk}$ on a 1-form basis which is dual to the system $\{\overline{e}_i\}$. Formally, the transformation between the 2-form and 2-vector is $\overline{A} = R^{op}\tilde{A}$, where $R^{op}$ is the index raising operator defined by the inverse metric tensor according to the component relation above. Conversely, we have $\tilde{A} = (R^{op})^{-1}\overline{A}$, which expressed in components is $A_{ij} = g_{il}g_{jk}A^{lk}$. The generalization of the maps to p-vectors and p-forms with arbitrary rank is obvious from this example.

*The outer product* (also called direct product) of the 1-form fields $\tilde{A}$ and $\tilde{B}$ is a (0/2)-tensor, denoted $\tilde{A}\otimes\tilde{B}$ (more detailed $\tilde{A}\otimes\tilde{B}(\quad,\quad)$). It is defined by $\tilde{A}\otimes\tilde{B}(\overline{c},\overline{d}) = \tilde{A}(\overline{c})\ \tilde{B}(\overline{d})$, where $\overline{c}$ and $\overline{d}$ are arbitrary vector fields. The generalization of the outer product to other types of tensors is straightforward. For example, the outer product of the vector fields $\partial/\partial x$ and $\partial/\partial y$ is a (2/0)-tensor field, denoted $\partial/\partial x\otimes\partial/\partial y$, which is defined as above but with 1-form arguments instead of vector arguments. If $\{\tilde{a}^i\}$ is a basis for 1-forms, then $\{\tilde{a}^i\otimes\tilde{a}^j\}$ is a basis for (0/2)-tensors, i.e. for an arbitrary (0/2)-tensor $\mathbf{D}$, we have $\mathbf{D} = D_{ij}\ \tilde{a}^i\otimes\tilde{a}^j$. If $\tilde{C}$ is a 2-form then $\tilde{C} = C_{ij}\ \tilde{a}^i\otimes\tilde{a}^j$, where $C_{ji} = -C_{ij}$.

*The wedge product* between 1-forms is defined by $\tilde{A}\wedge\tilde{B} = \tilde{A}\otimes\tilde{B} - \tilde{B}\otimes\tilde{A}$. For a 2-form $\tilde{C}$ we then have:

$$\tilde{C} = \frac{C_{ij}}{2}\ \tilde{a}^i\wedge\tilde{a}^j = C_{|ij|}\ \tilde{a}^i\wedge\tilde{a}^j, \tag{59}$$

where $\{\tilde{a}^i\}$ is a basis for 1-forms, and the bars around the indices mean that the summation is over indices $i < j$. Hence, $\{\tilde{a}^i\wedge\tilde{a}^j\}$ is a basis for 2-forms. In a similar way, $\{\overline{e}_i\wedge\overline{e}_j\}$ is a basis for 2-vectors. (The wedge product $\overline{e}_i\wedge\overline{e}_j = \overline{e}_i\otimes\overline{e}_j - \overline{e}_j\otimes\overline{e}_i$, and we have assumed that $\{\overline{e}_i\}$ is a vector basis.)

The wedge product can be extended to higher order forms and vectors. The principle is to write the wedge product as a sum of outer product terms where a term with an even permutation has a plus sign and a term with an odd permutation has a minus sign. If $\tilde{A}$, $\tilde{B}$ and $\tilde{C}$ are 1-forms the wedge product

$$\tilde{A}\wedge\tilde{B}\wedge\tilde{C} = \tilde{A}\otimes\tilde{B}\otimes\tilde{C} + \tilde{C}\otimes\tilde{A}\otimes\tilde{B} + \tilde{B}\otimes\tilde{C}\otimes\tilde{A} - \tilde{B}\otimes\tilde{A}\otimes\tilde{C} - \tilde{A}\otimes\tilde{C}\otimes\tilde{B} - \tilde{C}\otimes\tilde{B}\otimes\tilde{A},$$

is a 3-form. For a p-form $\tilde{D}$, with components $D_{i^1.....i^p} = \tilde{D}(\overline{e}_{i^1},......,\overline{e}_{i^p})$, where $i^\mu = 1, ......, n$, we have

$$\tilde{D} = \frac{D_{i^1.....i^p}}{p!}\ \tilde{a}^{i^1}\wedge.......\wedge\ \tilde{a}^{i^p} = D_{|i^1.....i^p|}\ \tilde{a}^{i^1}\wedge.......\wedge\ \tilde{a}^{i^p}. \tag{60}$$

$\{\tilde{a}^i\}$ is the dual basis to $\{\overline{e}_i\}$, and the bars around the indices mean that the summation is over indices $i^1 < i^2 < .......... < i^p$. If $p = n$, i.e. $p$ is equal to the dimension of the manifold, there is only one such index sequence, namely $123.....n$. A n-form therefore has only one independent component. If we use a coordinate basis, i.e. $\{\tilde{a}^i\} = \{\mathbf{d}x^i\}$, then an arbitrary p-form is expanded according to:

$$\tilde{D} = \frac{D_{i^1.....i^p}}{p!}\ \mathbf{d}x^{i^1}\wedge.......\wedge\ \mathbf{d}x^{i^p} = D_{|i^1.....i^p|}\ \mathbf{d}x^{i^1}\wedge.......\wedge\ \mathbf{d}x^{i^p}. \tag{61}$$

For p-vectors there are similar formulas; for example, for a p-vector $\overline{D}$ we have $\overline{D} = D^{|i^1.....i^p|}\ \overline{e}_{i^1}\wedge.......\wedge\overline{e}_{i^p}$, and if $\{\overline{e}_i\}$ is a coordinate vector basis $\{\partial/\partial x^i\}$, we have

$$\overline{D} = D^{|i^1.....i^p|}\ (\partial/\partial x^{i^1})\wedge.......\wedge(\partial/\partial x^{i^p}). \tag{62}$$

*A contraction formula.* We now give a formula - which applies in spaces with arbitrary dimension number - for the contraction between a vector $\overline{C}$ and a wedge product $\tilde{B}\wedge\tilde{A}$, where $\tilde{B}$ is a p-form and $\tilde{A}$ a form of any rank (the proof can be found in e.g. reference [8]):

$$(\tilde{B}\wedge\tilde{A})(\overline{C}) = \tilde{B}(\overline{C})\wedge\tilde{A} + (-1)^p\ \tilde{B}\wedge\tilde{A}(\overline{C}). \tag{63}$$

*A volume-form* is a n-form (where *n* is the dimension of the manifold) which can act on a set of *n* (small) vectors that form a generalized parallelepiped, resulting in a measure of the (generalized) volume. In three-dimensional Euclidean space we denote the volume-form by $\widetilde{\omega}$ and define it according to: $\widetilde{\omega} = \mathbf{d}x \wedge \mathbf{d}y \wedge \mathbf{d}z$. If we let $x^1 = x$, $x^2 = y$, $x^3 = z$, the components of the volume-form on the basis $\{\partial/\partial x^i\}$ satisfy

$$\omega_{ijk} = \widetilde{\omega}\Big(\frac{\partial}{\partial x^i}, \frac{\partial}{\partial x^j}, \frac{\partial}{\partial x^k}\Big) = \varepsilon_{ijk}. \tag{64}$$

$\varepsilon_{ijk}$ is the Levi-Civita symbol, which is 1 if *ijk* is an even permutation of 123, -1 if it is an odd permutation and 0 in other cases. The symbol $\varepsilon^{ijk}$ is also used and is defined in the same way as $\varepsilon_{ijk}$. Generalization to symbols in spaces of higher dimensions is evident from the definition here in three dimensions. In four-dimensional spacetime the volume-form is denoted $\widetilde{\Omega}$. For the Minkowski spacetime it is defined by

$$\widetilde{\Omega} = \mathbf{d}(ct) \wedge \mathbf{d}x \wedge \mathbf{d}y \wedge \mathbf{d}z, \tag{65}$$

where *c* is the velocity of light and the Cartesian coordinates *x, y, z* and corresponding vector and 1-form basis have been extended into four-dimensional spacetime. If we let $x^0 = ct$, $x^1 = x$, $x^2 = y$, $x^3 = z$, the components on the basis $\{\partial/\partial x^\alpha\}$ (where α = 0, 1, 2, 3) satisfy

$$\Omega_{\alpha\beta\gamma\delta} = \widetilde{\Omega}\Big(\frac{\partial}{\partial x^\alpha}, \frac{\partial}{\partial x^\beta}, \frac{\partial}{\partial x^\gamma}, \frac{\partial}{\partial x^\delta}\Big) = \varepsilon_{\alpha\beta\gamma\delta}. \tag{66}$$

*The 3-vector $\overline{\omega}$ and the 4-vector $\overline{\Omega}$.* Both of these n-vectors are defined from index raising operations on corresponding volume-forms. Thus $\overline{\omega} = R^{op}\ \widetilde{\omega}$ and $\overline{\Omega} = R^{op}\widetilde{\Omega}$. The components of $\overline{\omega}$ on the dual basis to the Cartesian vector basis $\{\partial/\partial x^i\}$ in (64) satisfy

$$\omega^{lmn} = g^{li}g^{mj}g^{nk}\omega_{ijk} = \varepsilon^{lmn}, \tag{67}$$

because the components of the metric tensor, and its inverse, are given by the unit matrix. The components of $\overline{\Omega}$ on the dual basis to $\{\partial/\partial x^\alpha\}$ in (66) satisfy

$$\Omega^{\alpha\beta\gamma\delta} = G^{\alpha\varepsilon}G^{\beta\eta}G^{\gamma\mu}G^{\delta\nu}\Omega_{\varepsilon\eta\mu\nu} = -\varepsilon^{\alpha\beta\gamma\delta}, \tag{68}$$

because the components of the metric tensor, and its inverse, for the Minkowski spacetime are given by the diagonal matrix $\mathrm{diag}(-1, 1, 1, 1)$.

*Dual operations. Dual fields.* The number of independent components for a p-form or a p-vector on a n-dimensional manifold is given by the expression $n!/(p!(n-p)!)$. For example, on a three-dimensional manifold, 1-forms, vectors (1-vectors), 2-forms and 2-vectors all have three independent components. Because p-forms, p-vectors, (n-p)-forms and (n-p)-vectors have the same number of independent components, one can find one-to-one mappings between the different objects. As was shown earlier, the metric tensor defines mappings between p-forms and p-vectors. If $\widetilde{p}$ is a given p-form then an associated p-vector is given by the relation $\overline{p} = R^{op}\widetilde{p}$, where $R^{op}$ is an index raising operator. In reverse order we have $\widetilde{p} = (R^{op})^{-1}\ \overline{p}$. Another type of operation which maps p-vectors into (n-p)-forms and p-forms into (n-p)-vectors is defined with the aid of the volume-form and the corresponding n-vector. For a given p-vector $\overline{p}$, with components $p^{j_1,\ldots,j_p}$ where $j_l = 1........n$, the dual (n-p)-form to $\overline{p}$, call it $\widetilde{q}$, is defined according to:

$$q_{i_1\ldots i_{(n-p)}} = \frac{1}{p!}\ \omega_{j_1\ldots j_p i_1\ldots i_{(n-p)}}\ p^{j_1,\ldots,j_p}, \tag{69}$$

where $\widetilde{\omega}$ here denotes the volume-form of a general manifold. The transformation is called a dual operation and is denoted by a star, we thus have $\widetilde{q} = {}^*\ \overline{p}$. If the p-vector is a 1-vector, i.e. a vector, the dual operation is identical to the contraction operation between the volume-form and the vector, i.e. ${}^*\ \overline{p} = \widetilde{\omega}(\overline{p},\ ,\ ...,\ )$.

If we start with a given p-form $\widetilde{p}$, with components $p_{j_1,\ldots,j_p}$, and let $\overline{\omega}$ denote the n-vector $R^{op}\widetilde{\omega}$, the relation

$$b^{i_1\ldots i_{(n-p)}} = \frac{1}{p!}\ \omega^{j_1\ldots j_p i_1\ldots i_{(n-p)}}\ p_{j_1,\ldots,j_p}, \tag{70}$$

defines a (n-p)-vector $\overline{b}$. Also this transformation, which maps p-forms into (n-p)-vectors, is in this paper called a dual operation and is denoted by a star, we thus have $\overline{b} = {}^*\ \widetilde{p}$. Consequently, we have two variants of dual operators, both denoted by a star. The definition of a dual operator depends on what object the operator acts on according to formulas (69) and (70). The result of two dual operations after another can be shown to satisfy the relations (see e.g. reference [8]):

$${}^{**}\widetilde{p} = \pm(-1)^{p(n-p)}\ \widetilde{p}, \quad {}^{**}\overline{p} = \pm(-1)^{p(n-p)}\ \overline{p}, \text{ i.e. } {}^{**} = \pm(-1)^{p(n-p)}\ I^{op},$$

where $I^{op}$ is the identity operator. The upper sign (plus) applies for cases where the manifold has definite metric, such as the Euclidean manifold. The lower sign (minus) applies for certain cases with indefinite metric, such as spacetime manifolds.

For three-dimensional Euclidean space: $** = 1^{\text{op}}$. For spacetime: $** = -(-1)^{p(4-p)} 1^{\text{op}}$. (71)

Assume $\bar{o}$ is a 0-vector field, i.e. a scalar field $\bar{o} = \varphi$, on a three-dimensional Euclidean manifold. From the definition of the dual operation in (69), it follows that the dual of the scalar field is equal to the scalar field times the volume-form, i.e. $*\varphi = \varphi\tilde{\omega}$. Therefore, because $** = 1^{\text{op}}$, we have

$$*(\varphi\tilde{\omega}) = \varphi. \tag{72}$$

*The exterior derivative*. The exterior derivative is an extension of the earlier defined gradient operator, which operates on functions, resulting in 1-form fields. The exterior derivative, here denoted $\mathbf{d}$, is defined to operate on form fields of arbitrary rank, resulting in form fields one rank higher. Thus, if $\tilde{a}$ is a p-form field then $\mathbf{d}\tilde{a}$ is a (p+1)-form field. In (73) below is the definition of the exterior derivative. $\tilde{b}$ and $\tilde{c}$ are q-form fields.

$$\mathbf{d}(\tilde{b} + \tilde{c}) = \mathbf{d}\tilde{b} + \mathbf{d}\tilde{c}, \ \mathbf{d}(\tilde{a} \wedge \tilde{b}) = \mathbf{d}\tilde{a} \wedge \tilde{b} + (-1)^p(\tilde{a} \wedge \mathbf{d}\tilde{b}), \ \mathbf{d}(\mathbf{d}\tilde{a}) = 0, \ \mathbf{d}f\left(\frac{\mathrm{d}}{\mathrm{d}\lambda}\right) = \frac{\mathrm{d}f}{\mathrm{d}\lambda}. \tag{73}$$

From these definitions we have $\mathbf{d}(f\mathbf{d}h) = \mathbf{d}f \wedge \mathbf{d}h$, where $f$ and $h$ are functions. It also follows that if a p-form $\tilde{a}$ is expressed in an arbitrary coordinate basis $\{\mathbf{d}x^i\}$, i.e.

$$\tilde{a} = \frac{a_{i_1 \dots i_p}}{p!} \mathbf{d}x^{i_1} \wedge \dots \wedge \mathbf{d}x^{i_p} = a_{|i_1 \dots i_p|} \mathbf{d}x^{i_1} \wedge \dots \wedge \mathbf{d}x^{i_p}, \text{ then}$$

$$\mathbf{d}\tilde{a} = \frac{1}{p!} \frac{\partial a_{i_1 \dots i_p}}{\partial x^k} \mathbf{d}x^k \wedge \mathbf{d}x^{i_1} \wedge \dots \wedge \mathbf{d}x^{i_p} = \frac{\partial a_{|i_1 \dots i_p|}}{\partial x^k} \mathbf{d}x^k \wedge \mathbf{d}x^{i_1} \wedge \dots \wedge \mathbf{d}x^{i_p}. \tag{74}$$

If $p = 2$ in (74), i.e. $\tilde{a}$ is a 2-form, it can be shown (e.g. by using the contraction formula (63) repeatedly) that the components of the 3-form $\mathbf{d}\tilde{a}$ on the vector basis $\{\partial/\partial x^i\}$ satisfy

$$\mathbf{d}\tilde{a}\left(\frac{\partial}{\partial x^i}, \frac{\partial}{\partial x^j}, \frac{\partial}{\partial x^k}\right) = (\mathbf{d}\tilde{a})_{ijk} = \frac{\partial a_{ij}}{\partial x^k} + \frac{\partial a_{ki}}{\partial x^j} + \frac{\partial a_{jk}}{\partial x^i} = a_{ij,k} + a_{ki,j} + a_{jk,i}, \tag{75}$$

where we have used the comma convention for the partial derivatives.

*Integration and Stokes' generalized theorem*. These topics are outlined in section 3B.